\documentclass[10pt,aps,prd,groupedaddress,showpacs,preprintnumbers,nofootinbib]{revtex4}
\usepackage{amssymb}
\usepackage{amsmath}
\usepackage{dcolumn}
\usepackage{rotating}

\usepackage{color}

\newcommand{\ba}{\begin{eqnarray}}
\newcommand{\ea}{\end{eqnarray}}
\newcommand{\be}{\begin{equation}}
\newcommand{\ee}{\end{equation}}
\newcommand{\bi}{\begin{itemize}}
\newcommand{\ei}{\end{itemize}}

\newcommand{\al}{\alpha}

\newcommand{\da}{\delta}
\newcommand{\la}{\lambda}

\newcommand{\sa}{\sigma}
\newcommand{\en}{\epsilon}

\newcommand{\La}{\Lambda}

\newcommand{\cF}{{\cal F}}
\newcommand{\cP}{{\cal P}}
\newcommand{\cQ}{{\cal Q}}
\newcommand{\cR}{{\cal R}}
\newcommand{\cO}{{\cal O}}

\newcommand{\cS}{{\cal S}}
\newcommand{\cT}{{\cal T}}

\newcommand{\w}{\widetilde}

\newcommand{\st}{\stackrel}

\newcommand{\p}{\partial}
\newcommand{\hp}{h^{\perp}}
\newcommand{\Ap}{A^{\perp}}
\newcommand{\n}{\nabla}

\newcommand{\ra}{\longrightarrow}

\newcommand{\LF}{\left(}
\newcommand{\RF}{\right)}
\newcommand{\LT}{\left[}
\newcommand{\RT}{\right]}

\newcommand{\gb}{\overline{g}}
\newcommand{\Rb}{\bar{R}}
\newcommand{\boxb}{\bar{\Box}}

\newcommand{\2}{\frac{1}{2}}

\newcommand{\4}{\frac{1}{4}}

\newcommand{\mx}{\mbox}
\newcommand{\mt}{\mathtt}

\newcommand{\where}{\mx{ where }}
\newcommand{\with}{\mx{ with }}

\newcommand{\ie}{{\it i.e.\ }}

\newcommand{\non}{\nonumber\\}

\newcommand{\Dc}{\mathcal{D}}
\newcommand{\Fc}{\mathcal{F}}

\newcommand{\pd}{\partial}
\newcommand{\cpd}{\nabla}
\newcommand{\D}{\nabla}

\newcommand{\const}{\text{const}}
\begin{document}

\title{Consistent Higher Derivative Gravitational theories with stable de Sitter and Anti-de Sitter Backgrounds}
\author{Tirthabir Biswas,}
\affiliation{Department of Physics, Loyola University, New Orleans 70118, USA}
\author{Alexey S. Koshelev,}
\affiliation{Departamento  de F\'isica and Centro  de  Matem\'atica  e
Aplica\c c\~oes,  Universidade  da  Beira  Interior,  6200  Covilh\~a,
Portugal}
\affiliation{Theoretische Natuurkunde, Vrije Universiteit Brussel, and The International Solvay Institutes, Pleinlaan 2, B-1050 Brussels, Belgium}
\affiliation{Steklov Mathematical Institute of RAS, Gubkina str. 8, 119991 Moscow, Russia }
\author{Anupam Mazumdar}
\affiliation{Consortium for Fundamental Physics, Lancaster University, Lancaster, LA1 4YB, United Kingdom}
\affiliation{Kapteyn Astronomical Institute, University of Groningen, 9700 AV Groningen, The Netherlands}

\date{\today}

\begin{abstract}
In this paper we provide the criteria for any generally covariant, parity preserving, and torsion free theory of gravity to possess a stable de Sitter (dS) or anti-de Sitter (AdS) background. By stability we mean the absence of tachyonic or ghost-like states in the perturbative spectrum that can lead to classical instabilities and violation of quantum unitarity. While we find that the usual suspects, the $F(R)$ and $F(G)$ theories,  can indeed possess consistent (A)dS backgrounds,  $G$ being the Gauss-Bonnet term, another interesting class of theories, string-inspired {\it infinite derivative  gravity}, can also be consistent around such curved vacuum solutions. Our study should not only be relevant for  quantum gravity and early universe cosmology involving ultraviolet physics, but also for modifications of gravity in the infra-red sector vying to replace dark energy.
\end{abstract}

\maketitle
\clearpage
\tableofcontents
\section{Introduction}
Einstein's General Relativity (GR) is an extremely successful theory in the infrared (IR), which matches a plethora of predictions and
observations, including various solar system tests and cosmological predictions~\cite{Will:2014kxa}. However, as it stands it has shortcomings in the ultraviolet (UV), it is incomplete classically as well as quantum mechanically; General Relativity admits blackhole and cosmological singularities, while the quantum loops render the theory non-renormalisable beyond
1-loop~\cite{'tHooft:1974bx}. In the case of a blackhole, at least the singularity
is covered by a horizon, but the cosmological singularity is ``naked''  where the energy density of the universe and all the curvatures blow up for physical time
$t\rightarrow 0$~\cite{Hawking:1973uf}. At the quantum level there have been many attempts to formulate a finite theory of gravity~\cite{Veltman:1975vx,dewittQG,DeWitt:2007mi} such as string theory (ST)~\cite{Polchinski:1998rr},  loop quantum gravity (LQG)~\cite{Ashtekar}, causal set~\cite{Henson:2006kf}, dynamical triangulation~\cite{Ambjorn:2012jv}, and asymptotic safety (AS)~\cite{Weinberg:1980gg} with varying degrees of success.  Intriguingly, most of these approaches to gravity have lead to non-local phenomena.

For instance in ST,  strings and branes are non-local objects with interactions spread over a region of space-time. Non-local structures also appear in noncommutative geometry and String Field Theory (SFT)~\cite{Witten:1985cc},  $p$-adic strings~\cite{Freund:1987kt}, zeta strings~\cite{Dragovich:2007wb}, and strings quantized on a random lattice~\cite{Douglas:1989ve,Biswas:2004qu}, for a review, see \cite{Siegel:1988yz}. LQG and causal set approaches are
primarily based on non-local Wilsonian operators, while non-locality in the form of an infinite set of derivatives have been discussed in the context of renormalization group arguments within the context of assymptotic safety~\cite{Krasnov}. It turns out that this appearance of an {\it  infinite series of higher-derivative} terms incorporating the non-locality, often  in the form of an {\it exponential kinetic} correction, is also a key feature  of many of the stringy constructions~\cite{Siegel:2003vt,Tseytlin:1995uq,Biswas:2005qr}. Thus, one of the main focus of our paper is to continue to investigate the consistency and viability of these infinite-derivative models and their implications for fundamental physics.

One of the typical challenges that any higher-derivative theory face is that they suffer from  Ostrogradsky
instabilities at a classical level, see~\cite{Woodard:2015zca}. They also appear while
canonically quantising the theory, see~ \cite{Eliezer:1989cr}. The Ostrogradsky argument relies on having a
highest "momentum" associated with the "highest derivative" in the theory in which the energy is seen
to be linear, as opposed to quadratic. This makes the energy of the system unbounded from below and signals the presence of
instability in the spectrum of the theory, which leads to lack of unitarity, predictability, and stability of the vacuum.

In gravity, a
classic example of a higher derivative theory that has ghosts is Stelle's $4th$ derivative theory of gravity~\cite{Stelle:1976gc},
which is renormalisable, but unfortunately contains a massive spin-2 ghost. In the path-integral approach, the presence of ghosts
can be identified from the  {\it extra poles/roots} that arise in the propagator with wrong sign residues.
As the Ostrogradski argument suggests, the issue of ghosts is hard to tame order by order, one is invariably  left with a highest momentum operator. In the language of the propagator, a finite fourth or higher order polynomial in momentum  will cause  trouble for the stability of the
action as in such a case it is easy to prove that the residue at one or more of the poles will inevitably have the wrong sign. In order to make sure that there exists
no extra poles in the propagator, one requires to modify the propagator by  an
{\it entire function}~\footnote{Often it is more convenient to impose restriction on the inverse momentum operator rather than the propagator. Since we do not want the propagators to have extra poles, this means the inverse propagator cannot have extra zeroes. This can be ensured if the inverse propagator is the exponential of an entire function which can never vanish in the finite complex plane.}, which contains no poles in the finite domain, and essential singularities only at the boundary, i.e. $\pm \infty$~\cite{Siegel:2003vt,Tseytlin:1995uq,Biswas:2005qr}. However, such a modification of a propagator also
demands that the theory must contain infinite derivatives. Since in this case there is no highest momentum associated with the highest derivative, the Ostrogradsky problem can be avoided and one is forced to work with the path integral formulation.
Perhaps not surprisingly, the stringy higher derivative modifications that we alluded before are precisely  of this form. While our paper will mainly focus on viability of such infinite derivative modifications in the context of gravity, our discussions and results are equally valid for most covariant higher derivative theories of gravity, including those that may be  relevant for inflation, for a review see~\cite{Mazumdar:2010sa}, or the dark energy problem~\cite{Langlois:2015cwa}.

In~\cite{Biswas:2011ar,Biswas:2013kla}, consistency of gravitational theories  around Minkowski space-time in $4$ space-time dimensions were investigated and concrete criteria were established to ensure the absence of any ghosts and tachyons in the perturbative spectrum. The analysis generalized the criteria that was found in~\cite{Biswas:2005qr} (see also~\cite{Biswas:2010zk,Biswas:2016etb} for robustness and perturbative stability of these models) for theories only containing the scalar curvature terms to include both the Ricci and Weyl tensor. In particular, the criteria reiterated the consistency of the widely popular $F(R)$ and $F(G)$ models where $R$ and $G$ are the Ricci scalar and Gauss-Bonnet scalars. It also corroborated the consistency of the class of IDG theories involving the Ricci scalar considered in~\cite{Biswas:2005qr} while demonstrating that these theories can also be viewed as an infinite derivative p-adic/SFT type scalar field nonminimally coupled to General Relativity, please see~\cite{overview} for a more detailed account. Most interestingly however, the authors, Biswas, Gerwick, Koivisto and Mazumdar (BGKM), also found a class of consistent infinite derivative gravitational (IDG) theories comprising up to quadratic terms in the Riemann tensor (not just the Ricci scalar) that contained no extra poles in the propagator other than the one corresponding to the massless graviton and no extra scalar degree of freedom.

In these new class of theories the only modification was in the form of a multiplicative {\it entire function} to the graviton propagator. In particular, in the ultraviolet (UV) the propagator could now become more convergent than the usual inverse square dependence of the momentum. Thus these theories can be thought of as ghost free infinite derivative extensions of Stelle's $4th$ derivative theory of gravity. In fact, it was shown in \cite{Talaganis:2014ida} that although softening of the propagator by an exponential inevitably implies an exponential enhancement
in the interactions in the UV, the superficial degree of divergence, $D$, which comes from a  combination vertex operators and propagators,  reads an encouraging, $D=1-L$, where $L$ is the number of loops. One can therefore hope that for $L>1$, the theory becomes super-renormalisable, similar results also hold for other forms of entire functions which are not exponentially suppressed~\cite{Tomboulis,Modesto}. The idea was tested and verified in a scalar-toy model of gravity up to 2-loops explicitly for an exponential propagator~\cite{Talaganis:2014ida}. For the same setup high energy scatterings of gravitons were also analysed, and it was found that vertices can overcome to the propagator
contributions at  finite loop levels to make the scattering diagram finite for a given external momenta~\cite{Talaganis:2016ovm}~\footnote{Infinite derivatives with Gaussian Kinetic term also has many applications in field theory~\cite{Biswas:2009nx}, and particle physics model building~\cite{Biswas:2014yia}.}.

While the quantum nature of IDG theories are encouraging, perhaps the most striking feature of IDG theories is their classical behaviour at the UV; the same propagator which softens quantum aspects of higher loops,
also seem to be able to avoid classical singularity for a point source - as long as the mass of the source, $m \lesssim \sqrt{M^2/M_p^2}$, where $M_p$ is the four dimensional Planck mass and $M$ is the scale of non-locality~\cite{Biswas:2011ar}. The classical avoidance of singularity was  tested vigorously in the linearised limit for both static~\cite{Edholm:2016hbt}, and time dependent cases~\cite{Frolov}. The avoidance of cosmological singularity has, in fact, been tested beyond linear level. Firstly, an  ansatz was recognized that resolved cosmological big bang singularity problem by replacing it with a big bounce that conjoins the expanding universe with a previous contraction~\cite{Biswas:2005qr}, see also~\cite{Biswas:2006bs,Koshelev:2012qn}. Secondly, these background solutions were perturbed on sub~\cite{Biswas:2012bp} and super-Hubble scales~\cite{Biswas:2010zk} to seek any unstable mode, but no instability has been observed yet, see also~\cite{Koshelev:2009ty,Koshelev:2010bf,Biswas:2015kha} for general features of perturbative evolutions that can also be applied to these bouncing scenarios and that further corroborates these findings. Finally, the avoidance of cosmological singularity has been tested at a non-linear level by understanding the geodesics of null rays to see whether they diverge or converge to test the focusing theorem of Penrose and Hawking~\cite{Penrose}. It was observed that IDG theories indeed give rise to de-focusing of null rays without violating any of the energy conditions required upon matter~\cite{Conroy:2014dja}, and see~\cite{Conroy:2016sac} for a detailed computation.

Last but not the least, an intriguing connection has  been established between the gravitational entropy, and the propagating degrees of freedom in the space-time. The gravitational entropy for ghost-free IDG does not get a contribution from the UV, but only from the Einstein-Hilbert action and follows strictly the area - law for entropy for a static spherically symmetric blackhole~\cite{Conroy:2015wfa}.

Given all the encouraging results that has emerged in the IDG theories, it stands to reason that we investigate the viability of these theories further. An obvious choice is to look at the consistency of other backgrounds that these theories may admit; after all a theory of quantum gravity should enable us to compute quantum amplitudes around any classical background, not just the Minkowski vacuum. The situation is similar to particle theories, while calculations around Minkowski space-time are the most important, field theories can be consistently expanded around solitonic backgrounds and provides sensible answers. Due to their simplicity as well as importance in cosmology and fundamental physics, looking at perturbations around de Sitter (dS) and Anti-de Sitter (AdS) space-times seemed the natural choice to make progress in this direction. On one hand, our hope is that the requirement of consistency around these curved backgrounds would provide us with additional constraint on IDG theories, and give us insights into  how a fundamental theory of gravity should look like. On the other hand, it is known that for several applications of gravity, ranging from testing gravity in our solar system to understanding cosmological phenomena, often one only needs to understand the dynamics of the relevant space-time background and linearised perturbations around it. We hope that not only our results for the dS background will aid inflationary or dark energy related cosmological model building efforts, but also  the techniques we have developed to study curved backgrounds will pave the way to investigate more nontrivial backgrounds, such as the Freedmann-Lema\^itre-Robertson-Walker (FLRW) space-times, and their perturbations.

In the first half of the paper we will review how to construct an ``equivalent action'' given any arbitrary covariant action of gravity that is parity preserving, torsion-free and admits a well-defined Minkowski limit. More details on this can be found in the companion book chapter~\cite{Biswas:2016etb}. The equivalent action  only consists of terms that are at most quadric in curvatures, but we will show that as far as the physics of the linearised fluctuations around dS/AdS are concerned, these actions are equivalent to the original higher derivative action that can potentially contain arbitrary high powers of curvatures. In the second half of the paper, we will vary the equivalent quadratic action of gravity around dS and AdS up to second order in fluctuations~\footnote{Previous studies have concentrated on finding the graviton propagator around dS and AdS backgrounds in
the context of Einstein-Hilbert action, see~\cite{Allen:1986ta,Antoniadis:1986sb,D'Hoker:1999jc,Miao:2011fc,Kahya:2011sy,Mora:2012zi}. Here we generalise to IDG.}.
We will decompose the $10$ metric components in $4$ space-time dimensions into the transverse and traceless spin-2 graviton field containing $5$ degrees of freedom, the $3$ transverse vector degrees of freedom, and  $2$ scalar degrees of freedom. There are, of course, $4$ gauge degrees of freedom, $3$ of which reduces the spin-2 field to the $2$ helicity states of the graviton, while the remaining gauge freedom is used to cancel the longitudinal vector mode reducing the vector  degrees of freedom to the two helicity states as well. This decomposition will explicitly demonstrate that just as in GR,  the vector and one of the scalars vanish from the action even for the higher derivative action. This can also be seen from the Bianchi identities that the field equations must satisfy. To summarise, we will be left with just the spin-2 graviton and $1$ scalar physical degree of freedom, indeed the latter is the familiar Brans-Dicke scalar that popularly arises in $F(R)$ theories. The final aim of this paper will be to write down explicitly the action for the graviton and the scalar mode in order to determine when these fields can propagate without encountering ghost like or tachyonic instabilities.

Our paper is organised as follows: In section 2, we are going to discuss how to construct a quadratic (in curvature) higher derivative action that is equivalent to an arbitrary covariant gravitational action as far as the physics around the constant curvature backgrounds (Minkowski, deSitter and anti-deSitter) are concerned, as long as the said action is torsionfree, parity-preserving and well defined in the Minkowski limit. We will provide illustrative examples of the construction for both local theories with a finite number of higher derivative terms and nonlocal theories with infinite derivative terms.  In section 3, we will compute the gravitational action keeping terms that are quadratic in fluctuations around (A)dS backgrounds, and then decompose this action into two parts corresponding to the physically surviving scalar and tensor modes of the metric. We will then proceed to obtain the consistency conditions for the theory to be free from ghost-like and tachyonic instabilities around (A)dS backgrounds in section 4. Apart from the usual local theories that are known to be consistent, we will see how consistent nonlocal IDG theories can also emerge. In particular, we will provide examples of IDG theories that provide consistent theories in the presence of an arbitrary cosmological constant, thus generalizing previous  constructions that provided viable theories only around Minkowski background. We will also provide three appendices: In appendix A we discuss various notations and identities, in appendix B we enumerate the commutation relations involving covariant derivatives that we will need in our computations, and in appendix C we provide the details of the cancellation of the vector and scalar modes in covariant gravitational actions.
\section{Equivalent Quadratic Actions around constant curvature backgrounds}
The main goal of this section is to start from the most general covariant, torsion-free, parity preserving quadratic action of gravity with a well defined Minkowski limit and obtain a simpler equivalent action that reproduces the same quadratic action for fluctuations around a constant curvature background,
such as dS and AdS. Here we will only sketch the steps, the detailed derivation can be found in  \cite{Biswas:2011ar} for a pure Minkowski space, and
for dS and AdS backgrounds in  \cite{Biswas:2016etb}. We impose the torsion free condition to ensure that the connection is not a separate field but is related to the metric so that we are only dealing with the degrees of freedom associated with the metric. Parity conservation imposes the condition that there cannot be any index contraction via the Levi-Cavita tensor and the only tensorial quantities that we are dealing with are the metric, the covariant derivatives and the curvatures. Finally, the fact that we demand our action to have a well defined Minkowski limit is to eliminate singular nonlocal terms. Our proof is often based on being able to expand our action in a Taylor series in curvatures, and expansions related to $1/R$ or $1/(\Box R)$ would involve non-analytic operators, see  \cite{Conroy:2014eja}. In future, one may be able to relax some of these conditions, but  with these restrictions, one can easily see that
the  most general action can be written in the form~\footnote{We are using $(-,+,+,+)$ signature of the metric.}:
\begin{equation}
S=\int d^4x\sqrt{-g}\LT \cP_0+\sum_i \cP_i\prod_I (\widehat{\cO}_{iI} \cQ_{iI})\RT\ ,
\label{general}
\end{equation}
where  $\cP_0,\cP_i,\cQ$ are quantities composed only of the Riemann and the metric tensor and $\widehat{\cO}$'s are  differential operators
{\it solely} constructed from covariant derivatives.  We are also only going to consider operators that are Taylor expandable {\it i.e. analytic}, so no terms such as $(1/\Box)R$ are present.

It is clear that the theory will admit dS, AdS and/or Minkowski vacuum solutions depending on its algebraic properties.
These backgrounds are of natural interest to cosmology, and AdS/CFT correspondence among other topics in high energy physics, for a review see~\cite{Aharony:1999ti}.
The AdS and dS backgrounds are maximally symmetric space-times, where we have
\begin{equation}
R=\bar R=\const,~~~~R_{\mu\nu}=\frac{\bar R}{4}\bar g_{\mu\nu},~~~~R^{\rho}_{\mu\sigma\nu}=\frac{\bar R}{12}(\delta^\rho_\sigma \bar g_{\mu\nu}-\delta^\rho_\nu \bar g_{\mu\sigma})\,.
\label{const-metrics}
\end{equation}
Hereafter ``$~\bar{}~$" designates the background quantity, and $\bar R=0$ in the above formulae yields the Minkowski
space-time~\footnote{The Greek indices run as: $\mu,\nu =0, 1, 2, 3$.}.
Now, in  \cite{Biswas:2016etb} it was shown that in order to study the perturbative properties of the action Eq.~(\ref{general}) around  dS,  AdS or Minkowski backgrounds, it is sufficient to look at a simpler equivalent action that just contains terms that are  quadratic in curvature, and potentially an infinite set of covariant derivatives. In other words, what we mean is that if the metric fluctuations are introduced, as
\begin{equation}
	g_{\mu\nu}= \bar g_{\mu\nu}+h_{\mu\nu}\ ,\label{deltagmunu}
\end{equation}
where $\bar g_{\mu\nu}$ is the background dS/AdS/Minkowski metric, then up to $\cO(h^2)$ the actions that one obtains from the original action and its equivalent action are identical.
The detailed procedure and rationale of reducing Eq.~(\ref{general}) to an equivalent quadratic action is described in  \cite{Biswas:2016etb}.
Here we summarise the algorithmic steps, and in each step we drop any terms which do not contribute to the variation of Eq.~(\ref{general}) around
constant curvature backgrounds given by Eq.~(\ref{const-metrics}). We now itemize the important steps:
\begin{enumerate}
\item The first stage is to reduce  the action Eq.~(\ref{general}) to an intermediate form
\begin{eqnarray}
S=\int d^4x\sqrt{-g}\LT \cP_0+\sum_{i=1}  \cP_i (\n \cR)(\widehat{\cO}_i \cR)\RT
\label{semi-final}
\end{eqnarray}
Here $\cR$ is a collective notation denoting either the Riemann tensor, Ricci tensor or the Ricci scalar. We proceed as follows:
\begin{enumerate}

\item Eliminate all the terms that are products of three or more terms that contain covariant derivatives. Note that we wish to keep only ${\cal O}(h^2)$ contributions.

\item If there are terms where the covariant derivative acts on a composite curvature term (\ie containing two or more Riemann tensors) then expand the terms
so that the covariant derivatives now only act on a single Riemann tensor.

\item Using integration by parts ensure that the relevant action is now of the form Eq.~(\ref{semi-final}). All the steps may have to be repeated
to reduce the action to the form (\ref{semi-final}).

\end{enumerate}
Let us illustrate the procedure with a couple of terms that  arises in 10th order gravity:
\begin{eqnarray}
S_a={1\over M^6}\int d^4x\sqrt{-g}\ \Box R \n_{\mu}R\n^{\mu}R\ ;~~~~\ S_b={1\over M^6}\int d^4x\sqrt{-g} R \n_{\mu}R\n^{\mu}R^2\ ; \nonumber \\
S_c={1\over M^6}\int d^4x\sqrt{-g}  R^{\mu\nu}\Box R_{\mu\nu} \Box R\ ,
\end{eqnarray}
where $\Box=g^{\mu\nu}\nabla_{\mu}\nabla_{\nu}$ is the d'Alembertian operator.
According to our prescription, the first term simply drops out as it is a product of three terms with derivatives. The second term  can be re-written as
$$S_b={2\over M^6}\int d^4x\sqrt{-g} R^2 (\n_{\mu}R)(\n^{\mu}R)$$
using simple product rule, while the third term can be re-expressed as
$$S_c=-{1\over M^6}\int d^4x\sqrt{-g}  R^{\mu\nu}(\n_\rho R)(\n^\rho\Box R_{\mu\nu}) +\dots\ ,$$
via integration by parts, where the dots indicate a term which has been dropped due to rule 1(a). We note that both these terms are now in the form Eq.~(\ref{semi-final}).
\item The second stage is to find the background solutions.

\begin{enumerate}
\item Identify the $\cP_0$ term by setting all the covariant derivative terms to zero. $\cP_0$ should just be a function of the curvatures.
\item Express the Riemann tensors in terms of the Ricci scalar, the traceless Ricci (TR) tensor
\be
S^{\mu}_{\nu}=R^{\mu}_{\nu} -\frac{1}{4}\delta^{\mu}_{\nu}R\,,
\ee
and the Weyl tensor
\be
C^{\rho}_{\mu\sigma\nu}= R^\mu_{\alpha\nu\beta} -\frac 1{2}(\delta^\mu_\nu
R_{\alpha\beta}-\delta^{\mu}_\beta
R_{\alpha\nu}+R^\mu_\nu{g}_{\alpha\beta}-R^\mu_\beta{g}_{\alpha\nu}
)+\frac R{6} (\delta^\mu_\nu {g}_{\alpha\beta}-\delta^{\mu}_\beta
{g}_{\alpha\nu})\,.
\ee
It is remarkably easy to work with  $S$ and $C$ tensors, since they  are both zero on the dS and AdS backgrounds.
We can now find the curvature $\bar R$ corresponding to the vacuum solution of interest, as a solution to equation
\begin{equation}
	\bar R=\left.\frac{2\cP_R}{\cP_{R}'}\right|_{R=\bar R}\where \cP_R\equiv \cP_0(R,C=0,S=0)\ ,
\label{backgroundKP}
\end{equation}
and the prime denotes derivative with respect to $R$.
\end{enumerate}
Since both $S_2$ and $S_3$ from the example above contain terms with derivatives they don't play any role in determining the background solution. For the purpose of illustration, let us consider the complete action for 2nd order gravity, or Stelle's theory~\cite{Stelle:1976gc},  instead:
\be
S_d=\int d^4x\sqrt{-g}\LT {M_p^2\over 2}R+c_{1,0} R^2 +c_{2,0} S^2+c_{3,0}C^2 -\La\RT \ .
\ee
All the terms contribute to $\cP_0$:
$$\cP_0={M_p^2\over 2}R+c_{1,0} R^2 +c_{2,0} S^2+c_{3,0}C^2 -\La\ ,$$
but the $\cP_R$ is just given by
\be
\cP_R={M_p^2\over 2}R+c_{1,0} R^2 -\La\ ,
\ee
so that $\bar{R}$ is given by the equation
\begin{equation}
\bar{R}={ M_p^2\bar{R}+2c_{1,0} \bar{R}^2 -2\La\over {M_p^2\over 2}+2c_{1,0} \bar{R} }~~~~~~~\Longrightarrow ~~~~~~~\bar{R}={4\La\over M_p^2}\ ,
\label{rlambdaalex}\end{equation}
a relation that we will come back to later.
\item The third stage is to find the equivalent action.

\begin{enumerate}
\item For the terms with derivatives, substitute the background values for curvatures Eq.~(\ref{const-metrics}) in the curvature functions,
$\cP_i$'s, which multiplies the derivative terms. Next commute the metric past the covariant derivatives, as follows
\begin{eqnarray}
&&\int d^4x\sqrt{-\gb} \overline{\cP}_i(\overline{\cR}) \da(\n \cR)\da(\widehat{\cO} \cR)\nonumber\\
&&=\int d^4x\sqrt{-\gb} \da(\n \cR)\da(\cP_i(\overline{g}_{\mu\nu}) \widehat{\cO} \cR)
=\int d^4x\sqrt{-\gb} \da(\n \cR)\da(\widehat{\cO'} \cR)\,.
\label{steps}
\end{eqnarray}
This is possible since the metric is annihilated by  covariant derivatives, i.e $\nabla_\lambda g^{\nu\mu}=0$.
At this point, the relevant action becomes
\begin{equation}
S=\int d^4x\sqrt{-g}\LT \cP_0+\sum (\n \cR)(\widehat{\cO}' \cR)\RT\,.
\label{intermediate}
\end{equation}
\item Next, use Bianchi identities to express the action in the following form
\begin{eqnarray}
S&=&\int d^4x\sqrt{-g}[\cP(\cR)+ R {\cal F}_{1}(\Box)R+S_{\mu\nu}
\Fc_2(\Box)S^{\mu\nu}+ C_{\mu\nu\la\sa}
\Fc_{3}(\Box)C^{\mu\nu\la\sa}]\,,\nonumber\\
\label{almost}
\end{eqnarray}
where the $\Fc_{i}$'s are of the form
\begin{equation}
\cF_{i}(\Box)=\sum_{n=1}^{\infty} c_{i,n}\Box^n\,.
	\label{Forms}
\end{equation}
The latter action is essentially the model considered in BGKM  \cite{Biswas:2011ar} with the
extra $\cP_0$ term~\footnote{One could rewrite  $\cF_{ i}(\Box)=\sum_{n=1}^{\infty} c_{i,n}\left({\Box}/{M^2}\right)^n$, where we have introduced a new
scale $M\leq M_p$ as the scale of non-locality in four dimensions.  Here $M_P=2.4\times 10^{18}$~GeV. To avoid cluttering of the formulae, we shall suppress writing $M$ explicitly, but where ever required for physical implications, we shall bring this to highlight its importance. }.

\item From the purely curvature term, $\cP_0$, the equivalent function, $\cP_{\mt{equiv}}$ can be written as
\begin{eqnarray}
\cP_{\mt{equiv}}=-\Lambda+\frac{M_P^2}2R+c_{R^2} R^2+c_{S^2}S_{\mu\nu}S^{\mu\nu}+c_{C^2}C_{\mu\nu\rho\sa}C^{\mu\nu\rho\sa}.
\end{eqnarray}
where
\ba
c_{R^2}&=&\frac12\cP''_R|_{R=\bar R},~\bar\cP_R=\cP_R(R=\bar R),~ M_P^2=\frac{4}{\bar R}(\bar\cP_R-c_{R^2}\bar R^2),~\Lambda=\bar\cP_R-c_{R^2}\bar R^2 \nonumber\\
c_{S^2}&=&\2\left({\p^2 \cP_0\over \p S_{\mu\nu}\p S^{\mu\nu}}\right)_{S=C=0,R=\Rb},\nonumber\\
c_{C^2}&=& \2\left({\p^2 \cP_0\over \p C_{\mu\nu\rho\sa}\p C^{\mu\nu\rho\sa}}\right)_{S=C=0,R=\Rb}\,.
\label{params}
\ea
As expected this results in
\begin{eqnarray}
M_P^2\bar R=4\Lambda\,,
\label{backgroundK2}
\end{eqnarray}
as it should be in accord with Eqs.~(\ref{backgroundKP}, \ref{rlambdaalex}).  Let us point out that there are several ``constant curvature'' space-times which has constant Ricci curvature, but they don't necessarily satisfy (\ref{const-metrics})~\footnote{Using gravitational field equations, one can prove that any metric which has a constant Ricci scalar, must also satisfy $\quad R_{\mu\nu}=\frac R4g_{\mu\nu}$ in our theory.}. Only maximally symmetric space-times, viz. dS, AdS and Minkowski space-times satisfy (\ref{const-metrics}) and therefore our analysis is only valid for these space-times.
\end{enumerate}

Let us illustrate this procedure for $S_b$ and $S_c$:
$$S_b={2\over M^6}\int d^4x\sqrt{-g} R^2 (\n_{\mu}R)(\n^{\mu}R)\rightarrow {2\over M^6}\int d^4x\sqrt{-g} \Rb^2 (\n_{\mu}R)(\n^{\mu}R)={2\Rb^2\over M^6}\int d^4x\sqrt{-g}R\Box R\ ,$$
while
$$S_c=-{1\over M^6}\int d^4x\sqrt{-g} R^{\mu\nu}(\n_\rho R)(\n^\rho\Box R_{\mu\nu})\rightarrow-{1\over 4M^6}\int d^4x\sqrt{-g} \Rb g^{\mu\nu}(\n_\rho R)(\n^\rho\Box R_{\mu\nu})$$
$$\rightarrow-{1\over 4M^6}\int d^4x\sqrt{-g} \Rb(\n_\rho R)(\n^\rho\Box R)=-{\Rb\over 4M^6}\int d^4x\sqrt{-g}  R\Box^2 R $$

\item Finally, we formulate the equivalent action as
\begin{equation}
S = \int d^4x\ \sqrt{-g}\left[\frac{M_P^2}2 R-\Lambda
+\frac{\lambda}2\left(R
\Fc_1(\Box)R+S_{\mu\nu}
\Fc_2(\Box)S^{\mu\nu}+ C_{\mu\nu\la\sa}
\Fc_{3}(\Box)C^{\mu\nu\la\sa}\right)
\right]\,.
\label{properaction}
\end{equation}
For  future convenience, we also write this action as
\begin{equation}
	S=S_{EH+\Lambda}+S_{R^2}+S_{S^2}+S_{C^2}\,.
	\label{properactionsch}
\end{equation}
The $\Fc$'s are of the form given by Eq.~(\ref{Forms}), except that the index $n$ now runs from 0 to $\infty$ instead of 1 to $\infty$, with the identification
\begin{equation}
c_{1,0}=c_{R^2},\quad\quad\quad c_{2,0}=c_{S^2},\quad\quad\quad c_{3,0}=c_{C^2}\,.
\label{params2}
\end{equation}
The dimensionless coupling $\lambda$ is introduced to control the higher derivative terms. For $\lambda \rightarrow 0$, we recover
Einstein-Hilbert action, while $M\rightarrow \infty$,  we recover purely a local action, quadratic in curvature.

To complete our illustration, if we had started with an action $S=S_a+S_b+S_c+S_d$, then the equivalent action would be given by Eq.~(\ref{properaction}), with
\ba
\cF_1(\Box)&=&c_{1,0}+ 2\Rb^2\Box-\4\Rb \Box^2=c_{1,0}+ 32{\La^2\over M^6M_p^4}\Box-{\La\over M^6M_p^2} \Box^2\,,\non
\cF_1(\Box)&=&c_{2,0}\,,\non
\cF_3(\Box)&=&c_{3,0}\ ,
\ea
while $\La$ and $M_p^2$, in this case remain unchanged. As expected, the coefficients in the equivalent theory indeed depend on the various parameters in the original action.

Finally, we note that the Gauss Bonnet scalar
\begin{equation}
	G=R^2-4R_{\mu\nu}^2+R_{\mu\nu\alpha\beta}^2=\frac16R^2-2S_{\mu\nu}^2+C_{\mu\nu\alpha\beta}^2\,,
	\label{GBscalar}
\end{equation}
being a topological invariant in four dimensions, allows us to set one of the coefficients among $c_{1,0},~c_{2,0},~c_{3,0}$ to zero, if we wanted to. Unfortunately, no such simplification is possible for higher derivative terms.
\end{enumerate}
\setcounter{equation}{0}
\section{Linearized non-local gravity on dS and AdS backgrounds}
In the previous section we have reviewed the process of obtaining an equivalent action starting from any covariant action as long as it is torsion free, parity-preserving and admits a well defined Minkowski limit. This essentially means that for the purpose of looking at dynamics of linear perturbations around dS, AdS or Minkowski background, it is sufficient to only focus on the simple equivalent action form given by Eq.~(\ref{properaction}). In this Section we will therefore start with the equivalent action Eq.~(\ref{properaction}) and obtain the $\cO(h^2)$ part of the action which is all that we will need to study perturbative consistency of these theories. Unfortunately, the calculations are rather technical, so we have summarised the main results in conclusion, in case the reader wants to skip these details.  Also, in our derivations, we very much rely on the formulae collected in Appendices, and wherever appropriate will direct the readers to the relevant sections in the Appendix.
\subsection{Quadratic Action}
\subsubsection{The Einstein-Hilbert term and $\Lambda$}
Let us start with the pure Einstein-Hilbert action with a cosmological term
\begin{equation}
	S_{EH+\Lambda} = \int d^4x\ \sqrt{-g}\left[\frac{M_P^2}2 R-\Lambda\right]\label{ehltext}\ .
\end{equation}
Its quadratic variation is obviously very well known, see for instance \cite{peter,chiba,solganik}, but for completeness sake we include them here. Using Eq.~(\ref{deltaEHfull}) from Appendix~\ref{ap_pert}, and assuming the background configuration Eq.~(\ref{const-metrics}), together with the relation Eq.~(\ref{backgroundK2}), we obtain the following quadratic variation for action Eq.~(\ref{ehltext}):
\ba
\delta^2 S_{EH+\Lambda}&=&\int
dx^4\sqrt{-\bar g}\frac{M_P^2}2\delta_0\,,\\
\delta_0&=&\left(\frac14 h_{\mu\nu}\bar \Box h^{\mu\nu}-\frac14h\bar\Box h+\frac12h\bar \D_\mu\bar \D_\rho h^{\mu\rho}+\frac12\bar \D_\mu h^{\mu\rho}\bar  \D_\nu h^\nu_\rho\right)-\frac1{48}\bar R\left(h^2+2h^\mu_\nu h^\nu_\mu\right)\,.
\label{d2GRadstext}
\ea

\subsubsection{The quadratic terms involving Ricci Scalar}
Again, utilising Eq.~(\ref{somedeltas}) from Appendix~\ref{ap_pert} for notations and actual computations, and using heavily that $\bar R=\const$, we obtain
\begin{equation}
\begin{split}
	\delta^2 S_{R^2}&=\frac\lambda2\int d^4x\sqrt{-\bar g}\left[2\left(\frac h2r+\frac12\left(\frac{h^2}8-\frac{h_{\mu\nu}h^{\mu\nu}} 4\right)\bar R+\delta^2(R)\right){c_{1,0}}\bar R+r\Fc_1(\bar\Box) r\right.\\
&\left.+\left(\frac h2\bar R+r\right)\delta(\Fc_1(\Box))\bar R+\bar R\delta^2(\Fc_1(\Box))\bar R+\frac h2\bar R(\Fc_1(\bar\Box)-{c_{1,0}})r+\bar R\delta(\Fc_1(\Box))r
\right]\,.
\end{split}
\label{drfr}
\end{equation}
where  we have already integrated by parts some terms. By inspection the terms in the first line can be rewritten, as
$$
\frac\lambda2\int d^4x\sqrt{-\bar g}\left[2{c_{1,0}}\bar R\delta_0+r\Fc_1(\boxb) r\right]\,.
$$
Now, turning to the second line of Eq.~(\ref{drfr}), we first notice that the first two terms are actually zero. This is because $\Rb$ is a constant and a scalar, and therefore annihilated by $\Box=\n^{\mu}\p_\mu$ as well as by $\delta\Box$ (see Eq.~(\ref{apdeltabox})). Also, for the last term, the variation of $\da \Fc_1(\Box)$ must appear in the $\Box$ appearing at the extreme left, otherwise, the term becomes a total derivative.
In other words, the last two terms in the second line of Eq.~(\ref{drfr}) can be combined as
$$
\frac\lambda2\sum_{n=1}^\infty{c_{1,n}}\bar R\int d^4x\sqrt{-\bar g}\left[\frac h2\bar\Box+\delta(\Box)\right]\bar \Box^{n-1}r\,,
$$
using the Taylor series representation for function $\Fc_1(\Box)$. Now, reciting explicitly Eq.~(\ref{apdeltabox})
$$\delta(\Box)\varphi=\left(-h^{\mu\nu}\cpd_\mu\pd_\nu-g^{\mu\nu}\gamma^\rho_{
\mu\nu } \pd_\rho\right)\varphi \,,$$ which is valid for any scalar field, $\varphi$, and integrating by parts  one can explicitly show that under the
integral, $\delta(\Box)\varphi$ is equivalent to $-\frac12(\bar\Box h)\varphi$, and therefore the two terms actually cancel. In other words, we have just proved
\begin{equation}
	\delta^2S_{R^2}=\frac\lambda2\int d^4x\sqrt{-\bar g}\left[2{c_{1,0}}\bar R\delta_0+r\Fc_1(\bar \Box) r\right]\,.
\label{Rterms}
\end{equation}

\subsubsection{Terms involving the TR and Weyl tensors \& the complete quadratic action}
Variations of the terms containing the Weyl or TR ternsors are extremely simple as both
these tensors are zero on constant curvature background, see Eq.~(\ref{const-metrics}), and they enter the action quadratically. This means
that the only terms which survive are
\begin{equation}
\begin{split}
\delta^2S_{S^2}=\frac\lambda2\int
d^4x\sqrt{-\bar g}\delta(S^\mu_{\nu})\Fc_2(\bar\Box)\delta(S^\nu_\mu)\quad \quad \quad\text{ and }\quad\quad\quad
\delta^2S_{C^2}=\frac\lambda2\int
d^4x\sqrt{-\bar g}\delta(C^{\mu\alpha}_{\phantom{\mu}\nu\beta}
)\Fc_3(\bar\Box)\delta(C_{\mu\alpha}^{\phantom{\mu\alpha}\nu\beta})
\end{split}
\label{SandC}
\end{equation}
Respective variations can be easily written in terms of $r$, $r^{\mu}_{\rho}$, $r^{\mu\nu}_{\phantom{\mu}\rho\sigma}$, these quantities being defined and computed in Eqs.~(\ref{somedeltas},\ref{extradeltas}), however one has to perform some algebraic manipulations to account properly all the contractions of the Kronecker symbols. A simplifying point is that
$r$, $r^{\mu}_{\rho}$, $r^{\mu\nu}_{\phantom{\mu}\rho\sigma}$ terms do not mix, thanks to the symmetry
properties of the Riemann tensor. We leave the explicit algebraic manipulations to the reader, and here just present the final result for the action containing quadratic fluctuations.

Summing all the individual contributions, Eqs.~(\ref{d2GRadstext}), (\ref{Rterms}), ({\ref{SandC}),  we get
\begin{equation}
\begin{split}
\delta^2 S=\int
dx^4\sqrt{-\bar g}&\left[\left(\frac{M_P^2}2+\lambda {c_{1,0}} \bar R\right)\delta_0+\frac\lambda2
\left(r\hat{\Fc}_1(\bar\Box)r+
r^\mu_\nu\hat{\Fc}_2(\bar\Box)
r^\nu_\mu+
r^{\mu\alpha}_{\phantom{\mu\alpha}\nu\beta}\hat{\Fc}_3(\bar\Box)
r_{\mu\alpha}^{\phantom{\mu\alpha}\nu\beta}
\right)
\right]
\end{split}
=s_0+s_1+s_2+s_3\,,
\label{d2action}
\end{equation}
where the following short notations are introduced:
\begin{eqnarray}
\hat{\Fc}_1(\bar\Box)&=& \Fc_1(\bar\Box)-\frac14\Fc_2(\bar\Box)+\frac13\Fc_3(\bar\Box)\,,\\
\hat{\Fc}_2(\bar\Box)&=& \Fc_2(\bar\Box)-2\Fc_3(\bar\Box)\,, \label{fbarf}\\
\hat{\Fc}_3(\bar\Box)&=& \Fc_3(\bar\Box)\ .
\end{eqnarray}
\subsection{Decoupling  Tensor, Vector, and  Scalar Modes }
Even though technically, we have derived the second order action, to understand the dynamical properties  we need to identify  the
physical excitations, or the correct propagating degrees of freedom. Now, any second rank tensor can be decomposed as, see for instance  \cite{D'Hoker:1999jc},
\begin{equation}
h_{\mu\nu}=\hp_{\mu\nu}+\bar\n_{\mu}\Ap_{\nu}+\bar\n_\nu\Ap_\mu+(\bar \n_{\mu}\bar\n_{\nu}-\4
\bar g_{\mu\nu}\bar \Box)B+\4 \bar g_{\mu\nu}h\,,\label{decomphabh}
\end{equation}
where the factor $4$ comes from dimensionality.
In $4$ dimensions, the metric tensor contains $10$ degrees of
freedom: $\hp_{\mu\nu}$, the transverse and traceless massless spin-two
graviton,
\begin{equation}
	\bar \n^{\mu}\hp_{\mu\nu}=\bar g^{\mu\nu}\hp_{\mu\nu}=0\,,\label{htt}
\end{equation}
represents $5$ degrees of freedom,  $\Ap_{\mu}$ the transverse vector field,
\begin{equation}
\bar\n^{\mu}\Ap_{\mu}=0\ ,
\end{equation}
accounts for $3$ degrees of freedom, and the two scalars, $B$ and $h$,
make up  the remaining two degrees of freedom. A priori, these fields
represent $6$ physical fields, since $3$ gauge degrees
reduce the spin two field to the $2$ spin-two helicity states of a graviton, and $1$ gauge freedom can be used to reduce the vector field to it's $2$ transversal spin-one helicity states as well. The aim of this section is to write down explicitly the action in terms of the tensor, vector and scalar components in order to
analyze their respective properties. To achieve
this we will use a variety of identities collected in the Appendix in order to
commute various derivative operators.

To begin with, we claim that once we directly substitute Eq.~(\ref{decomphabh})  in Eq.~(\ref{d2action}) all terms involving the vector field,
$A_\mu$, vanishes, and so does all the terms involving the $\bar \n_\mu\bar\n_\nu B$ piece. In other words, the quadratic action  only contains $\hp_{\mu\nu}$, and a single scalar field combination
\be
\phi\equiv \bar \Box B-h\ .
\ee
So, effectively Eq.~(\ref{decomphabh}) is reduced to
\begin{equation}
h_{\mu\nu}=\hp_{\mu\nu}-\4g_{\mu\nu}\phi\ .
\end{equation}
This result is identical to what happens in Einstein's gravity, but the algebraic computations needed to verify it for our general case are quite tedious, so we have briefly outlined it in the Appendices~\ref{apA} and \ref{apB}.

Next, we note that group representation theory dictates that at the linearised level, the tensor, vector and scalar
degrees should decouple from one another. Although, it is a well known fact for  pure GR, it may be not so transparent in a more general setting. The suspicion
comes from the presence of higher rank tensorial structures in the action. To
understand this deeper,  let us look at the GR terms in Eq.~(\ref{d2GRadstext}) first, as a warm-up exercise. Note that the tensor modes can in principal enter only in the very first and very last
terms in $\delta_0$. At any other place, they cancel due to the transverse and traceless
properties of $\hp_{\mu\nu}$ (see Eq.~(\ref{htt})). Thus if a mixing were to occur it can only be in the first or the last term in $\delta_0$. The relevant expression reads:
\begin{equation}
	\delta_{0,mix}=-\frac18 \hp_{\mu\nu}\bar\Box\bar g^{\mu\nu}\phi+\frac {\bar R}{48}{\hp}^\mu_\nu \delta^\nu_\mu\phi=0\,,
\end{equation}
 as $\hp_{\mu\nu}$ is traceless and $\bar \Box$ commutes with the metric tensor.

The higher derivative terms are less trivial. Essentially we need to analyze
expressions for $
r^{\mu\nu}_{\phantom{\mu}\rho\sigma}$.
Regarding the tensor modes, the structure of indices in the expression for
${r}^{\sigma\mu}_ {
\phantom{\mu}\nu\rho }$ and the transverse, traceless properties of
$\hp_{\mu\nu}$ suggest that one gains the expression for
${r}^{\sigma\mu}_
{ \phantom{\mu}\nu\rho }(\hp_{\mu\nu})$ by just replacing
$h_{\mu\nu}\to\hp_{\mu\nu}$. This trivial procedure yields
\begin{equation}
\begin{split}
{r}^{\sigma\mu}_ { \phantom{\mu}\nu\rho }(\hp_{\mu\nu})&=
\frac {\bar R}{24}(\delta^\mu_\nu {\hp}^\sigma_\rho-\delta^\mu_\rho
{\hp}^\sigma_\nu-\delta^\sigma_\nu
{\hp}^{\mu}_{\rho}+\delta^\sigma_\rho {\hp}^\mu_\nu)\\
&+\frac12
\left(\bar \cpd_\nu\bar \D^\mu
{\hp}^\sigma_\rho-\bar \D_\nu\bar\D^\sigma
{\hp}^\mu_\rho-\bar\cpd_\rho\bar\D^\mu {\hp}^\sigma_\nu+\bar\D_\rho\bar\D^\sigma {\hp}^\mu_\nu
\right)\,.
\end{split}
\label{tensorr4}
\end{equation}
The reason why nothing more can be simplified at this stage is because
$\hp_{\mu\nu}$'s are being acted by covariant derivatives with indices different
from those in $\hp_{\mu\nu}$. Therefore, no symmetry property can be utilised yet.

The scalar part of ${r}^{\sigma\mu}_ { \phantom{\mu}\nu\rho
}(\phi)$ allows some tinkering. The
simplification comes from the fact that no more than two derivatives appear and
they act on a scalar. In this case these derivatives commute, and we should not worry about
their order. The direct substitution gives
\begin{equation}
\begin{split}
{r}^{\sigma\mu}_ { \phantom{\mu}\nu\rho }(\phi)&=
\frac {\bar R}{48}(\delta^\sigma_\nu
\delta^{\mu}_{\rho}-\delta^\sigma_\rho \delta^\mu_\nu)\phi
-\frac18
\left(\bar\cpd_\nu\bar\D^\mu
\delta^\sigma_\rho-\bar\D_\nu\bar\D^\sigma
\delta^\mu_\rho-\bar\cpd_\rho\bar\D^\mu \delta^\sigma_\nu+\bar\D_\rho\bar\D^\sigma
\delta^\mu_\nu \right)\phi\,.
\end{split}
\label{scalarr4noDc}
\end{equation}
For  future convenience, we rewrite the latter expression as follows
\begin{equation}
\begin{split}
{r}^{\sigma\mu}_ { \phantom{\mu}\nu\rho }(\phi)&=
\frac18
\left(
\bar\Dc_\nu^\sigma
\delta^\mu_\rho+\bar\Dc_\rho^\mu
\delta^\sigma_\nu-\bar\Dc_\nu^\mu\delta^\sigma_\rho-\bar\Dc_\rho^\sigma
\delta^\mu_\nu \right)\phi
+\frac {3\bar\Box+\bar R}{48}(\delta^\sigma_\nu
\delta^{\mu}_{\rho}-\delta^\sigma_\rho \delta^\mu_\nu)\phi\,,
\end{split}
\label{scalarr4}
\end{equation}
where $$\bar\Dc_\nu^\mu=\bar\n_\nu\bar\n^\mu-\delta^\mu_\nu\frac{\bar\Box}4,\quad \bar\Dc^\mu_\mu=0,$$
i.e. it is a traceless operator, which will be extremely useful when evaluating the
action later.

So, the question is, whether any term survives in the following combination?
\begin{equation}
\begin{split}
\int
dx^4\sqrt{-g}&\left[{r}_{\sigma\mu}^{ \phantom{\sigma\mu}\nu\rho
}(\phi)\Fc(\bar\Box){r}^{\sigma\mu}_ { \phantom{\mu}\nu\rho
}(\hp_{\mu\nu})\right]\,.
\end{split}\label{mixsuspect}
\end{equation}
Schematically, four structures may arise
$$
\delta_\bullet^\bullet\delta_\bullet^\bullet\Fc\delta_\bullet^\bullet
{\hp}_\bullet^\bullet, \quad
\delta_\bullet^\bullet\delta_\bullet^\bullet\Fc\bar\n_\bullet\bar\n^\bullet
{\hp}_\bullet^\bullet, \quad
\delta_\bullet^\bullet\bar\n_\bullet\bar\n^\bullet\Fc\delta_\bullet^\bullet
{\hp}_\bullet^\bullet, \quad
\delta_\bullet^\bullet\bar\n_\bullet\bar\n^\bullet\Fc\bar\n_\bullet\bar\n^\bullet
{\hp}_\bullet^\bullet\,.
$$
Dots denote some indices, and each term is a scalar, \ie, indices are fully
contracted. Then,
we see that the first term goes away as finally you will have to contract the
indexes of $\hp_{\mu\nu}$ with a $\delta$. In the second term, in
order to avoid the appearance of the trace of $\hp_{\mu\nu}$, after the
$\delta$ contractions we must be left with $\bar\n^\mu\bar\n^\nu\hp_{\mu\nu}$, but this is zero as
$\hp_{\mu\nu}$ is transverse. Similarly, after the delta contractions the third term looks like $\bar\n^\mu\bar\n^\nu\Fc(\bar\Box)\hp_{\mu\nu}$. In Appendix~\ref{ap_commute} we have proved that $\bar\Box$ acting on the transverse and traceless symmetric second rank tensor gives again a transverse and traceless symmetric second rank tensor, and therefore $\Fc(\bar\Box)\hp_{\mu\nu}$ must be transverse and traceless ensuring that the third term also vanishes.

The fourth term generates four
possibilities upon contraction with the $\delta$-symbol
$$
\bar\n^\mu\bar\n^\nu\Fc\bar\n_\mu\bar\n_\nu{\hp}^\rho_\rho,\quad
\bar\n^\mu\bar\n^\nu\Fc\bar\n_\mu\bar\n_\rho{\hp}^\rho_\nu,\quad
\bar\n^\mu\bar\n^\nu\Fc\bar\Box{\hp}_{\mu\nu},\quad
\bar\n^\mu\bar\n^\nu\Fc\bar\n_\rho\bar\n_\nu{\hp}^\rho_\mu\,.
$$
It is easy to see that the first three terms vanish due to arguments similar to what we presented above. The last term is not that
obvious, however, but one finds that
$$
\bar\n_\rho\bar\n_\nu {\hp}^{\rho}_\mu=\frac {\bar R}3{\hp}_{\nu\mu}\,,
$$
which reduces it to the form of the third term.

To summarise, we have shown that in the quadratic action Eq.~(\ref{d2action}), only terms involving the transverse and traceless graviton field, ${\hp}_{\nu\mu}$, and a particular scalar field combination, $\phi$, survive. Further we have argued that the scalar and the tensor mode must decouple, and therefore it is sufficient to calculate the actions for scalar and tensor fields separately.
\subsection{Scalar modes}
Let us start with the scalar mode. Contracting Eq.~(\ref{scalarr4}) with the Kronecker delta over the first and
third indices, we immediately get
\begin{eqnarray}
{r}_{\mu\nu}(\phi)&=&\frac14\bar\Dc_{\mu\nu}\phi+\bar g_{\mu\nu}\LF\frac{3\bar\Box+\bar R}{16}\RF
\phi\ ,\label{first_nontrivial}
\end{eqnarray}
and then
\begin{eqnarray}
r(\phi)&=&\LF\frac{3\bar\Box+\bar R}4\RF\phi\ .
\end{eqnarray}
Also, quite a short computation is needed to get
\begin{eqnarray}
\delta_0(\phi)&=&-\frac1{32}\phi(3\bar\Box+\bar R)\phi\,.
\end{eqnarray}

We are now ready to look at the different terms in Eq.~(\ref{d2action}). The pure GR-part and the $r$-part require no simplifications, and one simply obtains
\ba
s_0+s_1\st{\phi}{\ra}  \int
dx^4\sqrt{-\bar g}\phi\left[-\frac1{32}\left(\frac{M_P^2}2+\lambda {c_{1,0}}\bar R\right) +{\lambda\over 32}\hat\Fc_1(\bar \Box)(3\bar \Box+\bar R)\right](3\bar \Box+\bar R)\phi\,.
\ea

Next, let us look at the term, $s_2$, involving $r_{\mu\nu}$.
By inspection, it is clear that there are three possible terms, terms containing two $ \bar\Dc^{\mu\nu}$'s, one $ \bar\Dc^{\mu\nu}$, and no $ \bar\Dc^{\mu\nu}$. The last one again doesn't require any simplification except for a trivial trace of the metric and the second term actually vanishes as $\bar\Dc_{\mu\nu}$ is traceless. Thus we are really left to evaluate terms such as
\begin{equation}
\int dx^4\sqrt{-\bar g}\ \left[\phi\bar \Dc_{\mu\nu}\Fc(\bar\Box)\bar\Dc^{\mu\nu}\phi\right]\,.\label{s2}
\end{equation}
As it will become progressively clear, in order to understand the dynamic properties of the fields, we will need to express the kinetic operators as functions of the $\Box$ operator. To achieve this  we have to commute covariant derivatives, which are on the left all the way to the right across an
infinite tower of d'Alembertians in the function $\Fc$. We can do this by utilising
the recursion property Eq.~(\ref{rec3}) derived in the Appendix~\ref{ap_commute}, which is appropriate since $\bar\Dc_{\mu\nu}$ is
traceless. Accordingly, we observe
\begin{equation}
  \bar\n_\nu\bar\Box^n
  \bar\Dc^{\mu\nu}\phi=\left(\bar\Box+\frac5{12} \bar R\right)^n \bar\n_\nu \bar\Dc^{\mu\nu}\phi=\left(\bar\Box+\frac5{12} \bar R\right)^n\left(\bar\Box\bar\n^\mu-\bar\n^\mu\frac{\bar\Box}
{ 4 } \right) \phi\,.
\label{s-recursion}
\end{equation}
Next, we utilise Eq.~(\ref{rec2}) along with Eq.~(\ref{rec3}) to obtain
\begin{eqnarray}
\bar\n_\mu\left(\bar\Box+\frac5{12}
\bar R\right)^n\left(\bar\Box\bar\n^\mu-\bar\n^\mu\frac{\bar\Box}
{ 4 } \right) \phi=\left(\bar\Box+\frac23
\bar R\right)^n\bar\n_\mu\left(\bar\Box\bar\n^\mu-\bar\n^\mu\frac{\bar\Box}
{ 4 } \right) \phi\nonumber \\
=\left(\bar\Box+\frac2{3}\bar R\right)^n\LF\frac{3\bar\Box+\bar R}4\RF\bar\Box\phi\,.
\end{eqnarray}
Note that the number $2/3$ arises as $5/12+1/4$. Returning to Eq.~(\ref{s2}), we can now
write down the cumulative expression
\begin{equation}
\begin{split}
\int dx^4\sqrt{-\bar g}\ \left[\phi\bar \Dc_{\mu\nu}\Fc(\bar\Box)\bar\Dc^{\mu\nu}\phi\right]
=\int dx^4\sqrt{-\bar g}\left[\phi
\Fc\left(\bar\Box+\frac2{3}
\bar R\right)\LF\frac{3\bar\Box+\bar R}4\RF\bar\Box
\phi\right]\,.
\end{split}
\label{st2}
\end{equation}
Finally, adding all the terms, we have
\be
s_2\st{\phi}{\ra} {\la\over 32}\int dx^4\sqrt{-\bar g}\ \phi\left[(3\bar\Box+\bar R)\hat\Fc_2(\bar\Box)+\bar\Box\hat\Fc_2\left(\bar\Box+\frac2{3}
\bar R\right)
\right]\LF\frac{3\bar\Box+\bar R}4\RF\phi\,.
\label{s2full}
\ee

Finally, for the $s_3$ part involving ${r}^{\sigma\mu}_ { \phantom{\mu}\nu\rho }$'s,
 we must carefully count all the non-vanishing products and respective coefficients. We
start by noticing that cross-products of terms with and without $\bar\Dc_{\mu\nu}$
again vanish as the trace of $\bar\Dc_{\mu\nu}$ (which is zero) would arise inevitable.
Then the simplest contribution is the one free of $\bar\Dc_{\mu\nu}$, and it reads
\begin{equation}
\phi\LF\frac {3\bar\Box+\bar R}{48}\RF\Fc_3(\bar\Box)
\LF\frac {3\bar \Box+\bar R}{48}\RF\phi
(\delta^\sigma_\nu
\delta^{\mu}_{\rho}-\delta^\sigma_\rho \delta^\mu_\nu)
(\delta_\sigma^\nu
\delta_{\mu}^{\rho}-\delta_\sigma^\rho \delta_\mu^\nu)=
\phi\LF\frac {3\bar\Box+\bar R}{2}\RF\Fc_3(\bar\Box)
\LF\frac {3\bar \Box+\bar R}{48}\RF\phi\,.
\end{equation}
Terms with $\bar \Dc_{\mu\nu}$ produce the following expression:
\begin{equation}
\frac18\phi
\left(
\bar \Dc_\nu^\sigma
\delta^\mu_\rho+\bar\Dc_\rho^\mu
\delta^\sigma_\nu-\bar\Dc_\nu^\mu\delta^\sigma_\rho-\bar\Dc_\rho^\sigma
\delta^\mu_\nu \right)\Fc_3(\bar\Box)
\frac18
\left(
\bar\Dc^\nu_\sigma
\delta_\mu^\rho+\bar\Dc^\rho_\mu
\delta_\sigma^\nu-\bar\Dc^\nu_\mu\delta_\sigma^\rho-\bar\Dc^\rho_\sigma
\delta_\mu^\nu \right)\phi=
\frac18\phi
\bar\Dc^\nu_\sigma\Fc_3(\bar\Box)
\bar\Dc_\nu^\sigma\phi
\end{equation}
This is a term of a type such as in Eq.~(\ref{st2}),
with the function $\hat\Fc_3$ inside, and with the coefficient $1/8$. Summing up the contributions, we get
\be
s_3\st{\phi}{\ra} {\la\over 32}\int dx^4\sqrt{-\bar g}\ \phi\LT\LF\frac {3\bar\Box+\bar R}{6}\RF\hat\Fc_3(\bar\Box)
+\2\bar\Box\hat\Fc_3\left(\bar\Box+\frac2{3}\bar R\right)\RT(3\bar\Box+\bar R)\phi\,.
\ee
Putting all the four terms $s_{0,1,2,3}$ together we now have the complete action for the scalar mode:
\be
\begin{split}
\label{d2properallRconstSCALARSbars}
&S_{0}=\frac1{32}\int dx^4\sqrt{-\bar g}\ \phi(3\bar \Box+\bar R)
\left\{ -\left(\frac{M_P^2}{2}+\lambda{c_{1,0}}\bar R\right)+\frac{\lambda}{2}\bigg[ 2\hat\Fc_1(\bar\Box)(3\bar \Box+\bar R)+\right.\\
&\left.\left.\frac{1}{2}\left(\hat{\Fc }
_2\left(\bar\Box+\frac{2}{3}
\bar R\right)\bar\Box+\hat{\Fc}_{2}\left(\bar\Box\right)(3\bar\Box+\bar R)\right)+
\left(\bar\Box{\hat\Fc}_3\left(\bar\Box+\frac{2}{3}\bar R\right)+
{\hat\Fc}_3\left(\bar \Box\right)\LF\frac{3\bar \Box+\bar R}{3}\RF \right)\right]\right\}\phi \,.
\end{split}
\ee
Notice that even though the functions $\Fc$ carry hats, the stand-alone ${c_{1,0}}$ term doesn't have a hat and this is the value corresponding to the function
$\Fc_1$. This latter action can actually be condensed slightly by reintroducing the functions $\Fc$'s without hats:
\begin{equation}
S_{0}=\frac1{32}\int
dx^4\sqrt{-\bar g}\ \phi(3\bar \Box+\bar R)\left\{-\left(\frac{M_P^2}2+\lambda {c_{1,0}}\bar R\right)
+\frac\lambda2\left[2\Fc_1(\bar \Box)(3\bar \Box+\bar R)+\frac1{2}{\Fc}_2\left(\bar \Box+\frac2 { 3 }\bar R\right)\bar\Box\right]
\right\}\phi\ .
\label{d2properallRconstSCALARSnobars}
\end{equation}
We note here that absence of the $\Fc_3$ function in the last formula is as to be expected. Indeed, if we restrict $h_{\mu\nu}$ to the $\phi$ part the   complete metric takes the form $g_{\mu\nu}=(1-\frac14\phi)\bar g_{\mu\nu}$. This is clearly a conformal scaling of the metric. The Weyl tensor of rank (1,3) is invariant under such a scaling and as a consequence a fully contracted square of the Weyl tensor is invariant as well. This implies that no contribution could arise from the Weyl tensor piece in the action, see Eq.~(\ref{properaction}).

\subsection{Tensor modes}
Let us now turn our attention to the tensorial terms.
Contracting Eq.~(\ref{tensorr4}) with the Kronecker delta over the first and
third indices and commuting the covariant derivatives,  we get
\begin{eqnarray}
	{r}_{\mu\nu}(\hp_{\nu\mu})&=&-\frac12 \LF \bar \Box-\frac {\bar R}6 \RF \hp_{\nu\mu}\,,
\end{eqnarray}
and consequently,
\begin{equation}
r(\hp_{\nu\mu})=0\ .
\end{equation}
We also have the well known result
\begin{equation}
	\delta_0(\hp_{\nu\mu})=\frac14\hp_{\nu\mu}\LF \bar \Box-\frac {\bar R}6 \RF{\hp}^{\nu\mu}\,,
\end{equation}
as this is the only term that appears in pure GR.

We  now have all the pieces to compute the action. The most challenging part was how
to compute the term containing $
r^{\mu\nu}_{\phantom{\mu}\rho\sigma}$. The trick which eventually allowed us to accomplish the task was
to roll-back to $\delta C^{\mu\nu}_{\phantom{\mu}\rho\sigma}$. To reduce
the clutter we will denote it as: $c^{\mu\nu}_{\phantom{\mu}\rho\sigma}$.
It can be obtained as a linear combination of $
r^{\mu\nu}_{\phantom{\mu}\rho\sigma}$, ${r}^\mu_{\nu}$ and $r$. It
enjoys all the symmetry (i.e. non-differential) properties of the Weyl tensor.
Presently, we are interested in tensor modes only. This means that we have
$r=0$. Although, generically a variation of a traceless tensor does not have to be
traceless, in our case the background is a conformally flat space time. For such space-times the Weyl tensor is zero. As a consequence $c^{\mu\nu}_{\phantom{\mu}\rho\sigma}$ is totally traceless similar to the Weyl tensor.

To see why the setup discussed here is so important, let us write down a generic term originating
from the part of the action Eq.~(\ref{d2action}) with the function $\hat\Fc_3$. It is of the form
\begin{equation}
\begin{split}
\int
dx^4\sqrt{-g}&\left[\hp_{\bullet\bullet}{\cO_L}_{\bullet\bullet}\Fc(\bar\Box)
\cO_R^{\bullet\bullet}
{\hp}^{
\bullet\bullet } \right ]\,.
\end{split}
\end{equation}
Here $\cO_{L,R}$ can be either a metric (delta-symbol), or two
covariant derivatives. Of course, the result is a scalar, all indices must be
contracted. Notice, however, that indices are always contracted across the
function $\Fc$, i.e. they are never contracted for tensor modes on one side of
the $\Fc$-factor (because this will generate either trace or transverse combination for $\hp_{\mu\nu}$, and both are zero). The most
tedious possibility is
$$
{\hp}_\bullet^\bullet\bar\n_\bullet\bar\n^\bullet\Fc\bar\n_\bullet\bar\n^\bullet
{\hp}_\bullet^\bullet\,,
$$
where indices can come in a large number of variations.

The biggest difficulty is to find a way of moving derivatives from the left
of $\Fc$ to the right of it. After rigorous computations, we  realised that we  needed to utilise recursion relations analogous to Eq.~(\ref{s-recursion}), which could only be obtained if tensors on the right have special symmetric properties.  $c^{\mu\nu}_{\phantom{\mu}\rho\sigma}$ became an appropriate choice, and indeed we obtained the following recursion relation (see (\ref{rec33})):
\begin{equation}
\begin{split}
\int
dx^4\sqrt{-\bar g}&\left[\hp_{\bullet\bullet}\bar\n_{\bullet}\bar\n_{\bullet}\Fc(\bar\Box)
c^{\bullet\bullet\bullet\bullet } \right]
=\int
dx^4\sqrt{-\bar g}\left[\hp_{\bullet\bullet}\bar\n_{\bullet}\Fc\left(\bar\Box+\frac {\bar R}4\right)
\bar \n_{\bullet}c^{\bullet\bullet\bullet\bullet } \right]\,.
\end{split}
\end{equation}
Here only two derivatives and not a metric can be on the left side of $\Fc$, as
everything else in this expression is totally traceless.

So, we want to first rewrite the action Eq.~(\ref{d2action}), as
\begin{equation}
\begin{split}
\delta^2 S=\int
dx^4\sqrt{-\bar g}&\left[\left(\frac{M_P^2}2+\lambda {c_{1,0}} \bar R\right)\delta_0+\frac\lambda2
\left(r\widetilde{\Fc}_1(\bar\Box)r+
r^\mu_\nu\widetilde{\Fc}_2(\bar\Box)
r^\nu_\mu+
c^{\mu\alpha}_{\phantom{\mu\alpha}\nu\beta}\w{\Fc}_3(\bar\Box){c}_{\mu\alpha}^{\phantom{\mu\alpha}\nu\beta}
\right)
\right]
=\w{s}_0+\w{s}_1+\w{s}_2+\w{s}_3\,,
\end{split}
\label{d2actionre}
\end{equation}
where the identification is obvious and the following short hand notations are introduced:
\begin{eqnarray}
\w{\Fc}_1(\Box)= \Fc_1(\bar\Box)-\frac14\Fc_2(\bar\Box)\,,~
\w{\Fc}_2(\bar\Box)= \Fc_2(\bar\Box)\,,\label{fbarfre}~
\w{\Fc}_3(\bar\Box)= \Fc_3(\bar\Box)\, .
\end{eqnarray}

The tensorial part from $\w{s}_3$ can now be computed as follows:
\begin{equation}
\begin{split}
\w{s}_3\st{h^\perp}{\ra}&\frac\lambda2\int
dx^4\sqrt{-\bar g}\left[
c^{\mu\alpha}_{\phantom{\mu\alpha}\nu\beta}(\hp_{\mu\nu})\Fc_3(\bar \Box)
c_{\mu\alpha}^{\phantom{\mu\alpha}\nu\beta}(\hp_{\mu\nu})
\right]\,,\\
=&\frac\lambda2\int
dx^4\sqrt{-\bar g}\left[
\frac12
\left(\bar \cpd_\nu\bar\D^\alpha
{\hp}^\mu_\beta-\bar\D_\nu\bar\D^\mu
{\hp}^\alpha_\beta-\bar\cpd_\beta\bar\D^\alpha {\hp}^\mu_\nu+\bar\D_\beta\bar\D^\mu
{\hp}^\alpha_\nu
\right)
\Fc_3(\bar\Box)
c_{\mu\alpha}^{\phantom{\mu\alpha}\nu\beta}(\hp_{\mu\nu})
\right]\,,\\
=&\frac\lambda2\int
dx^4\sqrt{-\bar g}\left[
2
{\hp}^\mu_\beta
\bar\D^\alpha\bar\cpd_\nu
\Fc_3(\bar\Box)
c_{\phantom{\nu}\mu\alpha}^{\nu\beta}(\hp_{\mu\nu})
\right]\,,\nonumber \\
=&\frac\lambda2\int
dx^4\sqrt{-\bar g}\left[
2
{\hp}^\mu_\beta
\bar\D^\alpha
\Fc_3\left(\bar\Box+\frac{\bar R}4\right)\bar\cpd_\nu
c_{\phantom{\nu}\mu\alpha}^{\nu\beta}(\hp_{\mu\nu})
\right]\,.
\end{split}
\end{equation}
Passing from the first to second line all other terms in the first $
c_{\mu\alpha}^{\phantom{\mu\alpha}\nu\beta}$ are dropped as they have at least
one $\delta$-symbol. The second transformation is solely due to the symmetry
properties of the tensor $
c_{\phantom{\nu}\mu\alpha}^{\nu\beta}$. Next we compute the rank-3 tensor
$\cpd_\nu
c_{\phantom{\nu}\mu\alpha}^{\nu\beta}$, and the result of quite a lengthy
computation is
\begin{equation}
\bar\cpd_\nu
c_{\phantom{\nu}\mu\alpha}^{\nu\beta}(\hp_{\mu\nu})=\frac14\left(\bar\Box-\frac{\bar R}4\right)\left(\bar\n_\alpha {\hp}^\beta_{\mu}-\bar\n_\mu {\hp}^\beta_\alpha\right)\,.
\end{equation}
One can check that the latter tensor satisfies all the properties required for
the use of recursion relation Eq.~(\ref{rec33}) apart from manifest symmetry with respect to
permutation of first two indices. This however is not necessary, as those
indexes are anyway contracted with a symmetric tensor $\hp_{\mu\nu}$ on the
left. Moving forward we get
\begin{equation}
\begin{split}
\w{s}_3\st{h^\perp}{\ra}&\int
dx^4\sqrt{-\bar g}\frac\lambda2\left[
\frac12
{\hp}^\mu_\beta
\bar\D^\alpha
\Fc_3\left(\bar\Box+\frac{\bar R}4\right)\left(\bar\Box-\frac{\bar R}4\right)\left(\bar\n_\alpha {\hp}^\beta_{\mu}-\bar\n_\mu {\hp}^\beta_\alpha\right)
\right]\,,\\
=&\int
dx^4\sqrt{-\bar g}\frac\lambda2\left[
\frac12
{\hp}^\mu_\beta
\Fc_3\left(\bar\Box+\frac{\bar R}3\right)\bar\D^\alpha\left(\bar\Box-\frac{\bar R}4\right)\left(\bar\n_\alpha {\hp}^\beta_{\mu}-\bar\n_\mu {\hp}^\beta_\alpha\right)
\right]\,,\\
=&\int
dx^4\sqrt{-\bar g}\frac\lambda2\left[
\frac12
{\hp}^\mu_\beta
\Fc_3\left(\bar\Box+\frac{\bar R}3\right)\left(\bar\Box-\frac{\bar R}6\right)\left(\bar\Box {\hp}^\beta_{\mu}-\bar\D^\alpha\bar\n_\mu {\hp}^\beta_\alpha\right)
\right]\,,\\
=&\int
dx^4\sqrt{-\bar g}\frac\lambda2\left[
\frac12
{\hp}^\mu_\beta
\Fc_3\left(\bar\Box+\frac{\bar R}3\right)\left(\bar\Box-\frac
{\bar R}6\right)\left(\bar\Box-\frac {\bar R}3\right) {\hp}^\beta_\mu
\right]\,.
\end{split}
\end{equation}
Since ${r}$ vanishes for the tensorial part we do not get any contribution from the $\w{s}_1$ part of the action, while the contribution from the  ${\widetilde s}_2$ part can be easily written down as ${r}_{\mu\nu}$  contains no covariant derivatives and therefore no commutations need be performed. Accordingly, we have
the final action for the tensor modes:
\begin{equation}
S_2=\frac14\int
dx^4\sqrt{-\bar g}\ \hp_{\nu\mu}\LF \bar \Box-\frac {\bar R}6
	\RF\left\{\frac{M_P^2}2+\lambda {c_{1,0}}
\bar R+
\frac\lambda2
\left[{\Fc}_2(\bar\Box)\LF \bar\Box-\frac {\bar R}6
\RF+2\Fc_3\left(\bar \Box+\frac{\bar R}3\right)
\left(\bar\Box-\frac{\bar R}3\right)\right]\right\}{\hp}^{\nu\mu}\,.
\label{d2properallRconstTENSORS}
\end{equation}
Please note that our final result contains functions $\Fc$ without tildes.
\setcounter{equation}{0}
\section{Physical Excitations \& Consistency Conditions}
\subsection{Canonical Action}
We are finally  ready to discuss the physics of the fluctuations, the main goal of our study.
At $\cO(h^2)$ the gravitational action has been neatly decomposed into a scalar and tensor part:
\begin{equation}
S_q=S_0+S_2\ ,
\end{equation}
with
\begin{eqnarray}
	S_2=\2\int dx^4\sqrt{-\bar g}~{\w{\hp}}^{\mu\nu} \LF \bar\Box-\frac {\bar R}6
	\RF \left\{1+{2\over M_p^2}\lambda {c_{1,0}}\bar R+
	{\lambda\over M_p^2}
	\left[\LF \bar \Box-\frac {\bar R}6
\RF{\Fc}_2(\bar \Box)+2\left(\bar \Box-\frac
{\bar R}3\right)\Fc_3\left(\bar\Box+\frac{\bar R}3\right)
\right]\right\}{\w{\hp}}_{\mu\nu}\,,\non
\end{eqnarray}
and
\begin{eqnarray}
S_0=-\2\int dx^4\sqrt{-\bar g}~\w{\phi}~\LF \bar \Box+{\bar R\over 3}\RF
\left\{1+{2\over M_p^2}\lambda{c_{1,0}}\bar R-{\lambda\over M_p^2}\left[2(3\bar \Box+\bar R)\Fc_1(\bar \Box)+
\frac1{2}\bar\Box{\Fc}
_2\left(\bar \Box+\frac2 { 3 }
\bar R\right)\right]
\right\}\w{\phi}\,,
\end{eqnarray}
where we have introduced canonical fields
\begin{eqnarray}
\w{\hp}_{\mu\nu}=\2M_p\hp_{\mu\nu}\,,~~~~~
\w{\phi}=\sqrt{3\over 32}M_p\phi\ .
\end{eqnarray}
In the Minkowski limit this yields the following spin-2 and spin-0 propagators:
\begin{eqnarray}
\Pi_{2}&=&{i\over p^2\left\{1-
{2p^2\over M_p^2}\left[{\Fc}_2(-p^2)+2\Fc_3\left(-p^2\right)
\right]\right\}}\,,\\
\Pi_0&=& {-i\over p^2\left\{1+{2p^2\over M_p^2}\left[6\Fc_1(-p^2)+
\frac1{2}{\Fc}
_2\left(-p^2\right)\right]
\right\}}\,,
\end{eqnarray}
where we have also chosen $\la=2$ to compare with the results obtained in~\cite{Biswas:2011ar}, and henceforth we will proceed with this identification. This agrees precisely~\footnote{To get a precise agreement with~\cite{Biswas:2011ar}, one also has to put $M_p=1$ and also realise that the scalar projection operator and the canonical field used here differs by a factor half, this is just a matter of convention.} with the results in~\cite{Biswas:2013cha}, once one realises that in the Ricci and  Riemann tensors were used to define $\cF$'s instead of the symmetric $S$-tensor and Weyl tensor that we use here. The translation is rather simple, the $\hat{\cF}_1$ and $\hat{\cF}_2$ defined in Eq.~(\ref{fbarf}) are nothing but the $\cF_1$ and $\cF_2$ in~\cite{Biswas:2013cha}, while the $\cF_3$ is unchanged.

One more useful check comes from comparing our results with the Gauss-Bonnet term, provided the curvature squared modification comes with
the form factors which must vanish, since the Gauss-Bonnet term is a topological invariant. This explicitly corresponds to fixing
$$
\Fc_1={c_{1,0}}=\frac16f_0,~~~\Fc_2={c_{2,0}}=-2f_0,~~~\Fc_3={c_{3,0}}=f_0\,.
$$
One can check that expressions in curly brackets in both $S_0$ and $S_2$ reduce to $1$, and one restores the pure  GR results.
\subsection{ghost free Condition}
The condition for absence of ghosts in our theory is equivalent to:
\begin{enumerate}
\item absence of new zeros in the spin-2 quadratic form as compared to the pure GR limit, and
\item presence of at most one more zero, say at $\Box=m^2$ with $m^2>0$, in the spin-0 quadratic form as compared to the pure GR limit.
An additional zero, if present, corresponds to the Brans-Dicke scalar mode usual in pure $F(R)$ gravity modifications.
\end{enumerate}
The above conditions mean that
\begin{equation}
	\cT(\bar R,\bar \Box)\equiv 1+{4\bar R\over M_p^2} {c_{1,0}}+{2\over M_p^2}
\left[\LF \bar\Box-\frac {\bar R}6
\RF{\Fc}_2(\bar\Box)+2\left(\bar\Box-\frac{\bar R}3\right)\Fc_3\left(\bar \Box+\frac{\bar R}3\right)\right]\,,
\end{equation}
should not have any zeroes, and
\begin{equation}
	\cS(\bar R,\bar \Box)\equiv 1+{4\bar R\over M_p^2}{c_{1,0}}-{2\over M_p^2}\left[2(3\bar\Box+\bar R)\Fc_1(\bar \Box)+\frac1{2}\bar \Box{\Fc}_2\left(\bar \Box+\frac2 { 3 }\bar R\right)\right]\,.
\end{equation}
can at least have a single zero. Note that since the zero of the scalar mode at $\Box=0$ has a wrong sign in the residue, the new zero, if present, is guaranteed to have the correct residue sign, and therefore will not be a ghost, while the constraint $m^2>0$ ensures that it is not tachyonic either.
Mathematically, we could express the two functions as
\begin{equation}
\cT(\bar R,\bar \Box)\equiv e^{\tau(\bar \Box)}\,,\label{cts1}
\end{equation}
and
\begin{equation}
\cS(\bar R,\bar \Box)\equiv \LF 1-{\bar\Box\over m^2}\RF^\epsilon e^{\sigma(\bar \Box)}\,,\label{cts2}
\end{equation}
with $\tau$ and $\sigma$ being  entire functions, $\epsilon=0,1$ and $m$ real. $\epsilon=0$ corresponds to no extra scalar mode and $\epsilon=1$ corresponds to the Brans-Dicke scalar.
\subsection{Illustrative Examples}
In this subsection we want to provide a few simple examples of gravitational models, which are consistent around dS or AdS backgrounds. In particular, we will focus on the cases when only one of the three functions, $\Fc_{1,2,3}$'s, is non-zero. In this process, we will also extend the IDG model with only quadratic curvature terms that has been shown previously~\cite{Biswas:2011ar} to consistently modify the graviton propagator, and ameliorate some of the UV problems of GR without introducing any new degrees of freedom, ghosts or otherwise. By explicit construction, we will show how by including nonlinear (in curvature) terms in the action it is possible to have a gravitational theory that is not only consistent around the Minkowski background ($\La=0$), but also the curved dS ($\La>0$) or AdS ($\La<0$) backgrounds. It will become evident that nonlinear terms are necessary in order to achieve this, and corroborates the idea that requiring quantum the theory of gravity to be consistent around all possible backgrounds may be a powerful way to constrain the modifications to GR.
\subsubsection{$\cF_1\neq0$, but $\Fc_{2}=\Fc_{3}=0$}
In this case, we have manifestly no extra poles in the spin-2 propagator. However, we must ensure that
\begin{equation}
	1+{4\bar R\over M_p^2} {c_{1,0}}=1+{16\La\over M_p^4} {c_{1,0}}>0\,,
\label{zeroeth}
\end{equation}
and also guarantee, that
\begin{equation}
	\cS(\bar R,\bar \Box)= 1+{4\bar R\over M_p^2}{c_{1,0}}-{4\over M_p^2}(3\bar\Box+\bar R)\Fc_1(\bar\Box)=\LF 1-{\bar\Box\over m^2}\RF^\epsilon e^{\sigma(\bar\Box)}\,.
\end{equation}
We note that for the constant terms to match on both sides we must have $\sa(0)=0$. Then, the function $\Fc_1$ has to be of the form
\be
\Fc_1(\bar\Box)={1+{4\bar R\over M_p^2}{c_{1,0}}-\LF 1-{\bar\Box\over m^2}\RF^\epsilon e^{\sigma(\bar\Box)}\over {4\Rb\over M_p^2}\LF 1+{3\bar\Box\over \Rb}\RF}={1+{16\La\over M_p^4}{c_{1,0}}-\LF 1-{\bar\Box\over m^2}\RF^\epsilon e^{\sigma(\bar\Box)}\over {16\La\over M_p^4}\LF 1+{3M_p^2\bar\Box\over 4\La}\RF}\,.
\label{cf1}
\ee
A couple of comments are now in order: Firstly, as one looks at the last expression, it is clear that for any set of parameters, $\La,~ c_{1,0},~m,~\en$, and analytic function $\sa(\Box)$, we have a $\cF_1$ that gives rise to consistent fluctuations around a specific dS or AdS background, as long as Eq.~(\ref{zeroeth}) is satisfied, and $\sa(0)=0$. However, by inspection it is also clear that  the function, $\cF_1$, depends on the value of $\La$, and it is not possible therefore for a single $\cF_1$ function to be simultaneously consistent for two different $\La$'s. In other words, if we are starting from a model with only quadratic curvature terms, it is impossible for the theory to be consistent for arbitrary values of the cosmological constant. On the other hand, one may imagine that a truly consistent theory of gravity should be consistent in presence of any arbitrary stress-energy tensor, and in particular, for any value of the cosmological constant. The form of $\cF_1$ in terms of $\Rb$ actually suggests a simple way out of this problem. If we instead allow for nonlinear terms in our original gravitational action, such as an action of the form
\begin{equation}
S = \int d^4x\ \sqrt{-g}\left[\frac{M_P^2}2 R-\Lambda
+R\Fc_1(\Box,R)R\right]\with \Fc_1(\Box,R)={1+{4R\over M_p^2}{c_{1,0}}-\LF 1-{\Box\over m^2}\RF^\epsilon e^{\sigma(\Box)}\over {4R\over M_p^2}\LF 1+{3\Box\over R}\RF}\ ,
\label{1action}
\ee
then such an action will be consistent for any arbitrary value of $\La$, and in fact, around a given background ($\La$) the equivalent $\cF_1$ will be precisely given by Eq.~(\ref{cf1}).

Secondly, let us point out that this is a theory, where essentially we have GR coupled to a  scalar field theory. If $\sa$ identically vanishes, then $\en=1$ reproduces Brans-Dicke theory. For instance, if $\sa(\Box)=-\Box$, and $\en=0$, then we have a p-adic type scalar field coupled to gravity. Such systems have been studied in details in the context of cosmology~\cite{Barnaby:2006hi}. On the other hand, if we have  $\sa(\Box)=-\Box$, and $\en=1$, this corresponds to an SFT type tachyonic field coupled to gravity whose cosmological implications have also been studied in previous literature~\cite{Koshelev:2009ty}. Furthermore, this also provides an extension of Starobinsky's original model of inflation~\cite{Starobinsky} to seek an UV completion, see~\cite{Biswas:2013dry} and detailed perturbation analysis in~\cite{Craps:2014wga,Koshelev:2016xqb}.
\subsubsection{$\cF_2\neq0$, but $\Fc_{1}=\Fc_{3}=0$}

In this case  we get two constraints on the same function $\Fc_2$:
\begin{equation}
\cT(\bar R,\bar \Box)\equiv 1+{2\over M_p^2}
\LF \bar \Box-\frac {\bar R}6
\RF{\Fc}_2(\bar \Box)= e^{\tau(\bar \Box)}
\end{equation}
\begin{equation}
\cS(\bar R,\bar \Box)\equiv 1-{2\over M_p^2}
\frac1{2}\bar\Box{\Fc}_2\left(\bar\Box+\frac2 { 3 }\bar R\right)=(\bar\Box-m^2)^\epsilon e^{\sigma(\bar\Box)}
\end{equation}
and the only solution is the trivial case $\Fc_2=0$. In other words,  we get back to GR.


\subsubsection{$\cF_3\neq0$, but $\Fc_{1}=\Fc_{2}=0$, the case of only gravitons}

A particularly interesting and illuminating case, which has been previously discussed in the context of black hole and big bang singularity is when no scalar degrees of freedom are present, and the presence of infinite covariant derivatives only modify the graviton propagator without introducing any new states. This is ensured by demanding that $\cS$ is just a constant, which, in particular, can be achieved by setting $\Fc_{1}=\Fc_{2}=0$.
We are then left with a graviton quadratic form, that reads
\begin{equation}\w{\hp}_{\mu\nu}
	{\LF -\bar \Box+\frac {\bar R}6
	\RF\left[1-{4\over M_P^2}\left(-\bar\Box+\frac {\bar R}3\right)\Fc_3\left(\bar \Box+\frac{\bar R}3\right)\right]}\w{\hp}^{\mu\nu}
\end{equation}
In order to illustrate; how one can obtain a graviton propagator involving an infinite set of higher derivatives, let us consider the ``simplest'' case where the modified quadratic form is as follows:
\begin{equation}\w{\hp}_{\mu\nu}
	{\LF -\bar \Box+\frac {\bar R}6
\RF e^{\al\LF -\bar \Box+{\bar R\over 6}\RF}}\w{\hp}^{\mu\nu}\ .
\label{nonlocal-graviton}
\end{equation}
This provides an exponential suppression at high momentum, which has been found to ameliorate the black hole and big bang singularities, but does not alter the Newtonian limit as the residue at $\bar \Box=\bar R/6$ remains unaltered. It is easy to obtain the form of  $\cF_3$ function which gives rise to an inverse propagator, such as Eq.~(\ref{nonlocal-graviton}):
\begin{equation}
	\cF_3(\Box)= {M_p^2\over 4}\LT{e^{-\al\LF \Box -{ R\over 2}\RF}-1\over {\Box}-{2R\over 3}}\RT\,.
\end{equation}
Again, just as our construction in Eq.~(\ref{1action}), such a nonlinear function ensures that the model remains consistent in the presence of any cosmological constant, the coefficients adjust appropriately when perturbed around any given dS/AdS/Minkowski background so that no new poles are introduced.

The above function does have a pole at $\Box=2 R/3$, which is perfectly acceptable as the propagators are still well defined. However, it is also possible to construct analytic functions. For instance,
\begin{equation}
	\cF_3(\Box)= {M_p^2\over 4}\LT{e^{\al\LF \Box -{ R\over 2}\RF\LF  \Box -{2 R\over 3}\RF}-1\over {\Box}-{2 R\over 3}}\RT\,,
\end{equation}
is an analytic function yielding the spin-2 quadratic form
\begin{equation}\w{\hp}_{\mu\nu}
	{\LF -\bar \Box+\frac {\bar R}6
\RF e^{\al\LF -\bar \Box+{\bar R\over 6}\RF\LF -\bar \Box+{\bar R\over 3}\RF}}\w{\hp}_{\mu\nu}\ ,
\end{equation}
that again only contains the graviton pole and an exponential suppression at high momenta.
\section{Conclusions}

In this paper we have provided an algorithm to construct the most general parity invariant and torsion free covariant action of gravity
that is consistent around constant curvature backgrounds, as long as the action has a well defined Minkowski limit. In particular, we have studied dS and
AdS backgrounds. Our analysis smoothly reduces to the Minkowski space-time limit which was studied before by BGKM in~\cite{Biswas:2011ar}. Our prescription is generic, and is equally applicable for both UV and IR higher derivative modifications that has found various cosmological and stringy applications. We also checked our analysis against some well known cases; for instance, when the scale of non-locality, $M\rightarrow \infty$, the class of consistent gravitational actions reduced to the popular local models of $4$ derivative gravity, the $F(R)$ and $F(G)$ theories. We paid special attention to the Gauss-Bonnet action as it provided us with some nontrivial checks on our derivations.

We found that the most general action can indeed contain {\it infinite} covariant derivatives, along with Ricci scalar, Ricci tensor and Weyl/Riemann terms. The infinite derivatives can be expressed in terms of {\it form factors}, whose forms can be now constrained by demanding that the action is perturbatively
 {\it ghost} and {\it tachyon free} around constant curvature backgrounds. In order to verify this, it was sufficient to perturb the gravitational action
 up to order ${\cal O}( h^2)$ around dS and AdS backgrounds. We  computed explicitly the second order variation of Einstein-Hilbert term, and
the higher order terms comprising the form factors. For pedagogical reasons, we have provided details of our conventions in Appendix \ref{Append-1}, properties of
 maximally symmetric space-times in \ref{Append-2}, and perturbations in \ref{ap_pert}. Some very useful and powerful identities around constant curvature backgrounds were obtained in \ref{ap_commute}.

 In $4$ dimensions, in order to obtain the true propagating degrees of freedom in space-time, we had to decompose the metric tensor
 into its degrees of freedom corresponding to transverse-traceless, $h_{\mu\nu}^{\perp}$, a transverse vector field, $A_{\mu}^{\perp}$, and $2$
 scalars, $B,~h$. After performing a systematic decomposition  we were able to show that the only viable propagating degrees of freedom are the transverse-traceless
 spin-2 field, $h_{\mu\nu}^{\perp}$, and a scalar combination which gives rise to spin-0 component, $\phi = \bar\Box B-h$. The transverse vector component, $A_{\mu}^{\perp}$, and
 $\nabla_\mu\nabla_{\nu} B$ vanish identically, leaving the massless graviton and possibly a Brans-Dicke scalar to propagate around dS, AdS or Minkowski backgrounds. The details of the latter computation can be found in appendices,  \ref{apA}, \ref{apB}.

In order to make the graviton propagator perturbatively {\it ghost free}  in dS and AdS backgrounds, one  has to make sure that the second variation of spin-2 component does not introduce any new pole in the propagator, and
must recover the pure GR limit at low energies. Similarly, the spin-0 component can allow at most one extra pole, say at mass $m^{2}> 0$ in its propagator. The latter condition is necessary to ensure that an additional pole representing the Brans-Dicke scalar, if present as in $F(R)$ gravity modifications, is not a tachyon. If both spin-2 and spin-0 propagators do not have any extra poles, then both can be expressed in terms of an exponential of an {\it entire function}, which does not introduce any pole except the essential singularity at the boundary corresponding to the  UV limit, $\Box \rightarrow \infty$. In the IR limit, when $\Box \rightarrow 0$, the second variation of both spin-2 and spin-0 components
recover the GR limit in dS and AdS backgrounds.

We have illustrated that we can recover various limits from the most generic IDG action. In the limit when ${\cF}_2=0,~\cF_3=0,~\cF_1\neq 0$, the theory effectively reduces to scalar-tensor theory around
dS and AdS backgrounds. When $\cF_1=0,~\cF_3=0$, the {\it ghost free} condition enforces a simple solution where $\cF_2=0$, too, thereby reducing to pure GR. In the last scenario, when $\cF_1=0,~\cF_2=0,~\cF_3\neq 0$, the  theory space reduces to pure spin-2 excitations, and no physical spin-0 mode.

Our analysis indeed opens up new avenues for higher derivative theories of gravity, including IDG, which can be made consistent at both classical and at quantum level around dS and AdS backgrounds.
This should have important implications for both cosmology and for AdS/CFT correspondence. In the dS case, it provides the possibility to realise a Big Bounce~\cite{Biswas:2004qu}, avoiding Big Bang singularity problem. This scenario also
presents the first viable UV generalisation of  Starobinsky inflation~\cite{Starobinsky}, and also provides an interesting connection between the graviton degrees of freedom propagating in spacetime with the avoidance of focusing the null congruences in a time dependent background~\cite{Conroy:2016sac}. 

Furthermore, our analysis also provides strong link to stable  ghost and tachyon free modifications of GR in the IR, in the context of dark energy problems. In this context it will be rather useful if one can  extend the class of actions we have analyzed to include terms that do not necessarily have a well defined Minkowski limit as such actions have been discussed considerably in the literature to address the the dark energy problem.

In the  case of AdS, a consistent IDG provides an ideal platform to study connection between gravity in the UV, and the corresponding CFT in the boundary.  After all, presence of IDG is inevitable in
closed string theory, in terms of $\alpha'$ corrections. At present, computing all order $\alpha'$ correction in AdS background in the closed string sector is indeed challenging, and although our analysis
does not involve supersymmetry, we believe that constructing stable and consistent theory of gravity around AdS will  help us in ascertaining how such an action can be derived from closed string field theory.

Finally, we would like to emphasize that several of the computations and strategies that we developed in analyzing perturbations around dS/AdS should carry over to more nontrivial backgrounds such as FLRW or Black hole space-times. As mentioned before, most of the applications and tests of GR involves encoding the physics around certain highly symmetric background space-times and small perturbations around them. Thus, we believe that our analysis could go a long way in making progress towards analysing such important physical space-times for very general covariant theories of gravity.

\section{Acknowledgements}

AK is supported by the FCT Portugal fellowship SFRH/BPD/105212/2014 and in part by FCT Portugal grant UID/MAT/00212/2013 and RFBR grant 14-01-00707. The work of A.M. is supported in part by the Lancaster-Manchester-Sheffield Consortium for Fundamental Physics under STFC grant ST/L000520/1.

\appendix
\section{Notations}\label{Append-1}

\subsection{General Backgrounds}

Here we introduce the notations used at the background level.

The metric is
$$g_{\mu\nu}=(-,+,+,+,\dots),\quad g_{\mu\nu}g^{\mu\nu}=D=4.$$
The dimension is always 4 and $4$-dimensional indexes are small Greek letters. The metric
connection (Cristoffel symbols) is
$$
\Gamma_{\mu\nu}^\rho=\frac12g^{\rho\sigma}(\pd_\mu g_{\nu\sigma}+\pd_\nu
g_{\mu\sigma}-\pd_\sigma g_{\mu\nu})\,,
$$
The covariant derivative is $\nabla_\mu$ and acts as
$$
\cpd_\mu F^{.\alpha.}_{.
\beta.}=\pd_\mu F^{.\alpha.}_{.
\beta.}+\Gamma^\alpha_{\mu\chi}F^{.\chi.}_{.
\beta.}-\Gamma^\chi_{\mu\beta}F^{.\alpha.}_{.
\chi.}
$$
It follows that $\nabla_\rho g^{\rho\nu}\equiv0$.
The Riemann tensor, curvatures ,and the Einstein tensor are defined as
$$
R^\sigma_{\mu\nu\rho}=\pd_\nu\Gamma^\sigma_{\mu\rho}
-\pd_\rho\Gamma^\sigma_{\mu\nu}+\Gamma^\sigma_{\chi\nu}\Gamma^\chi_{\mu\rho}
-\Gamma^\sigma_{\chi\rho}\Gamma^\chi_{\mu\nu}
,\quad R_{\mu\rho}=R^\sigma_{\mu\sigma\rho},\quad R=R^\mu_\mu,\quad
G_{\mu\nu}=R_{\mu\nu}-\frac12Rg_{\mu\nu}.$$
The symmetry properties are
\begin{equation*}
\begin{split}
 R_{\mu\nu\rho\sigma}=-R_{\mu\nu\sigma\rho}=-R_{\nu\mu\rho\sigma
} &=R_ { \rho\sigma\mu\nu},\quad
R_{\mu\nu\rho\sigma}+R_{\mu\sigma\nu\rho}+R_{\mu\rho\sigma\nu}=0,\quad
R_{\mu\nu}=R_{\nu\mu}\,.
\end{split}
\end{equation*}
The commutator of covariant derivatives is
$$
[\cpd_\mu,\cpd_\nu]A_\rho=R^\chi_{\rho\nu\mu}A_\chi\,,
$$
The  Bianchi identity is given by:
$$
\cpd_\lambda R_{\mu\nu\rho\sigma}+\cpd_\sigma R_{\mu\nu\lambda\rho}+\cpd_\rho
R_{\mu\nu\sigma\lambda}=0\,.
$$
It implies
\begin{equation*}
  \begin{split}
    \cpd_\mu R^\mu_\nu=\frac12\pd_\nu R,\quad \cpd_\nu\cpd_\mu
R^{\mu\nu}=\frac12\Box R,\quad
    \cpd_\lambda R_{\nu\sigma}-\cpd_\sigma R_{\nu\lambda}+\cpd^\mu
R_{\mu\nu\sigma\lambda}=0,\quad \cpd_\mu G^\mu_\nu=0\,.
  \end{split}
\end{equation*}
The d'Alambertian (the box) is defined as
$
\Box=g^{\mu\nu}\cpd_\mu\cpd_\nu
$,
and acts in a fully covariant way. Another useful operator is
$$
\Dc_{\mu\nu}=\n_\mu\n_\nu-\frac14g_{\mu\nu}\Box\,.
$$
It is traceless and this simplifies certain computations.
The traceless analog of the Einstein tensor is given by:
$$
S_{\mu\nu}=R_{\mu\nu}-\frac1DRg_{\mu\nu},\quad \quad\quad S_\mu^\mu=0\,.
$$
The Weyl tensor follows from  the Ricci decomposition, and this is given by (in $D$ space-time dimensions)
\begin{equation*}
 C^\mu_{\alpha\nu\beta}= R^\mu_{\alpha\nu\beta} -\frac 1{D-2}(\delta^\mu_\nu
R_{\alpha\beta}-\delta^{\mu}_\beta
R_{\alpha\nu}+R^\mu_\nu{g}_{\alpha\beta}-R^\mu_\beta{g}_{\alpha\nu}
)+\frac R{(D-2)(D-1)} (\delta^\mu_\nu {g}_{\alpha\beta}-\delta^{\mu}_\beta
{g}_{\alpha\nu})\,.
 \end{equation*}
The Weyl tensor has all the symmetry properties of the
Riemann tensor and also it is absolutely traceless,
i.e.
$
C^\mu_{\alpha\mu\beta}=0
$.
Moreover this rank (1,3) tensor is invariant under the conformal scaling of the metric. The latter implies that the
Weyl tensor is zero on conformally flat manifolds, i.e.  on the manifolds where the metric can be brought to the form $ds^2=a(x)^2\eta_{\mu\nu}dx^\mu dx^\nu$, with
$\eta_{\mu\nu}$ being the Minkowski metric with the same signature as the original one.
The following quadratic relations always hold:
$$
S_{\mu\nu}^2=R_{\mu\nu}^2-\frac14R^2,\quad
C_{\mu\nu\alpha\beta}^2=R_{\mu\nu\alpha\beta}^2-2R_{\mu\nu}^2+\frac13R^2.
$$
The Gauss-Bonnet term can be written as
$$
G=R^2-4R_{\mu\nu}^2+R_{\mu\nu\alpha\beta}^2=\frac16R^2-2S_{\mu\nu}^2+C_{
\mu\nu\alpha\beta}^2.
$$
\subsection{Maximally symmetric space-times}\label{Append-2}

By definitions maximally symmetric space-times have $\frac12D(D+1)$ linearly independent
Killing vectors. This translates into the fact that
\begin{equation}
R^\sigma_{\mu\nu\rho}=\frac R{D(D-1)}(\delta^\sigma_\nu
g_{\mu\rho}-\delta^\sigma_\rho g_{\mu\nu})\,.\label{maxsym}
\end{equation}
In general $R$ does not have to be constant. One however can prove
using the Bianchi identities that for $D>2$ this form of the Riemann
tensor implies, $R=\const$. Consequently, in such space-times
$$
\cpd_\lambda R^\sigma_{\mu\nu\rho}=0
$$
Also, one readily sees
\begin{equation}
S_{\mu\nu}=0,\quad \quad \quad C^\sigma_{\mu\nu\rho}=0\,.  \label{zeroSC}
\end{equation}
Both AdS and dS are  maximally symmetric (i.e. it satisfies
Eq.~(\ref{maxsym})). $R=\const>0$ for dS, and  $R=\const<0$ AdS. Minkowski
has $R^\sigma_{\mu\nu\rho}=0$, and can be seen as the $R\to 0$ limit of the
(A)dS space-time.
\subsection{Perturbations}\label{ap_pert}
Here we introduce notations and quantities relevant for computing the second
variation of the action around an (A)dS background. The metric variation is
$$g_{\mu\nu}= \bar g_{\mu\nu}+h_{\mu\nu}.$$
Bar's are used to designate the background quantities.
The following relations are relevant for perturbation analysis in this paper
\begin{equation}
\begin{split}
g^{\mu\nu}\to& \bar g^{\mu\nu}-h^{\mu\nu},\quad\quad \sqrt{-g}\to\sqrt{-\bar g}\left(1+\frac
h2+\frac{h^2}8-\frac{h^\mu_\nu h^\nu_\mu}4\right),\quad \quad h=\bar g^{\mu\nu}h_{\mu\nu}\,,\\\\
\Gamma^\mu_{\nu\rho}\to&\bar \Gamma^\mu_{\nu\rho}+\gamma^\mu_{\nu\rho},~\gamma^\mu_{
\nu\rho}=\frac12(\bar \cpd_\nu h^\mu_\rho+\bar \cpd_\rho h^\mu_\nu-\bar \cpd^\mu
h_{\nu\rho}),~\quad\quad \gamma_{\mu\rho}^\rho=\frac12\pd_\mu h\,,\\\\
R^\sigma_{\mu\nu\rho}\to&
\bar R^\sigma_{\mu\nu\rho}+\widetilde{r}^\sigma_{\mu\nu\rho},~\widetilde{r}^\sigma_{\mu\nu\rho}
=\bar \cpd_\nu\gamma^\sigma_{\mu\rho}-\bar \cpd_\rho\gamma^\sigma_{\mu\nu}\,,
\\\\
R^{\sigma\mu}_{\phantom{\mu}\nu\rho}\to&
\bar R^{\sigma\mu}_{\phantom{\mu}\nu\rho}+{r}^{\sigma\mu}_{\phantom{\mu}\nu\rho
} , ~\quad\quad {r}^{\sigma\mu}_ { \phantom{\mu}\nu\rho }=
-h^{\mu\tau}\bar R^\sigma_{\tau\nu\rho}+\bar g^{\mu\tau}\widetilde{r}^\sigma_{\tau\nu\rho}\,,
\\\\
R_{\mu\rho}\to&
\bar R_{\mu\rho}+\widetilde{r}_{\mu\rho}+\gamma^\sigma_{\chi\sigma}\gamma^\chi_{\mu\rho}
-\gamma^\sigma_{\chi\rho}\gamma^\chi_{\sigma\mu},~\quad \widetilde{r}_{
\mu\rho }
=\bar \cpd_\nu\gamma^\nu_{\mu\rho}-\bar \cpd_\rho\gamma^\nu_{\mu\nu}=\frac12(\bar \D_\nu\bar \D_\mu
h^\nu_\rho+\bar \D_\nu\bar \D_\rho h^\nu_\mu-\bar \Box h_{\mu\rho}-\bar \D_\rho\pd_\mu h)\,,
\\\\
R^\mu_{\rho}\to&
\bar R^\mu_{\rho}+{r}^\mu_{\rho},~\quad {r}^\mu_{\rho}
=-h^{\mu\sigma}\bar R_{\sigma\rho}+\bar g^{\mu\sigma}\widetilde{r}_{\sigma\rho}\,,
\\\\
R\to&
\bar R+r,~\quad \quad r
=-h^{\mu\rho}\bar R_{\mu\rho}+\bar g^{\mu\rho}(\bar \cpd_\nu\gamma^\nu_{\mu\rho}
-\bar \cpd_\rho\gamma^\nu_ { \mu\nu
})=(-\bar R_{\mu\nu}+\bar \D_\mu\bar \D_\nu-\bar g_{\mu\nu}\bar \Box)h^{\mu\nu}\,.
\end{split}\label{somedeltas}
\end{equation}
Here the arrow means that the RHS is equal to the LHS up to higher order corrections. The order of expansion is either linear or quadratic in above expressions.
One can show that other (higher order in $h$) terms do not contribute to the quadratic variation of the action. They either generate more than quadratic corrections
to the quadratic action variation, or become total derivatives.
Note that
\begin{equation*}
  \delta F^\mu = \delta(g^{\mu\nu} F_\nu)=-h^{\mu\nu}F_\nu+\bar g^{\mu\nu}\delta
F_{\nu}\neq \bar g^{\mu\nu}\delta F_\nu\,.
\end{equation*}
The two quantities which are often actually used are
\begin{equation}
\begin{split}
	{r}^\mu_{\rho}&=-\frac {\bar R}4h^{\mu}_\rho+\frac12(\bar \D_\nu\bar \D^\mu
h^\nu_\rho+\bar \D_\nu\bar \D_\rho h^{\nu\mu}-\bar \Box h^\mu_{\rho}-\bar \D_\rho\pd^\mu h)\,,\\
{r}^{\sigma\mu}_ { \phantom{\mu}\nu\rho }&=
\frac {\bar R}{24}(\delta^\mu_\nu h^\sigma_\rho-\delta^\mu_\rho
h^\sigma_\nu-\delta^\sigma_\nu
h^{\mu}_{\rho}+\delta^\sigma_\rho h^\mu_\nu)+\frac12
\left(\bar \cpd_\nu\bar \D^\mu
h^\sigma_\rho-\bar \D_\nu\bar \D^\sigma
h^\mu_\rho-\bar \cpd_\rho\bar \D^\mu h^\sigma_\nu+\bar \D_\rho\bar \D^\sigma h^\mu_\nu \right)\,.
\end{split}\label{extradeltas}
\end{equation}
Note that,
$$
r=\delta^\rho_\sigma
r^{\sigma}_{\rho}\,,~~~~~~
r^{\sigma}_{\rho}=\delta^{\nu}_{\mu}r^{\mu\sigma}_{\phantom{\mu}\nu\rho}\,.
$$
Upon acting on scalars, the variation of the $\Box$ operator is
\begin{equation}
\delta(\Box)\varphi=\left(-h^{\mu\nu}\bar \cpd_\mu\pd_\nu-\bar g^{\mu\nu}\gamma^\rho_{
\mu\nu } \pd_\rho\right)\varphi\,.
\label{apdeltabox}\end{equation}
As a warm up exercise, given pure Einstein-Hilbert action with a cosmological term, i.e.
\begin{equation*}
S_0 = \int d^4x\ \sqrt{-g}\left[\frac{M_P^2}2 R-\Lambda\right]\,,
\end{equation*}
one obtains a second variation around any arbitrary background, as
\begin{equation}
\begin{split}
\delta^2 S_{0}=\int
dx^4\sqrt{-\bar g}\frac{M_P^2}2&\left[\left(\frac14 h_{\mu\nu}\bar \Box
h^{\mu\nu}-\frac14h\bar \Box h+\frac12h\bar \D_\mu\bar \D_\rho
h^{\mu\rho}+\frac12\bar \D_\mu h^{\mu\rho}\bar \D_\nu h^\nu_\rho\right)\right.+\\
&+(hh^{\mu\nu}-2h^\mu_\sigma h^{\sigma\nu})\left(\frac18
\bar g_{\mu\nu}\bar R-\frac14 \bar g_{\mu\nu}\frac\Lambda{M_P^2}-\frac12 \bar R_{\mu\nu}\right)-\\
&-\left.\left(\frac12\bar R_{\sigma\nu}h^\sigma_\rho
h^{\nu\rho}+\frac12\bar R^\sigma_{\rho\nu\mu}h^\mu_\sigma h^{\nu\rho}\right)\right]\,.
\end{split}
\label{deltaEHfull}
\end{equation}
This result can be checked against, e.g., \cite{Christensen:1979iy}. Note that the indices of perturbed quantities are raised and lowered by the background metric.
\section{Commutation relations}\label{ap_commute}
Here we collect important commutation relations which are used in transforming
the expressions. All the expressions in this section are written for maximally symmetric space-times in mind. Also, all the derivative operators, metric, and curvatures take their background values, and we simply omit bars for clarity.

Given an arbitrary scalar $\varphi$, we have
\begin{eqnarray}
  \n_\mu\n_\alpha \varphi&=&\n_\alpha\n_\mu \varphi \,,\\
  \n_\mu\n_\alpha\n_\beta \varphi&=&\n_\alpha\n_\beta\n_\mu \varphi+\frac
R{12}(g_{\beta\mu}\n_\alpha-g_{\alpha\beta}\n_\mu)\varphi\,,\\
  \n_\mu\Box \varphi&=&\left(\Box-\frac R4\right)\n_\mu \varphi\,.
  \label{rec1}
\end{eqnarray}
Given an arbitrary vector $t_\mu$, we have
\begin{eqnarray}
  \n_\mu\n_\alpha t^\mu&=&\n_\alpha\n_\mu t^\mu+\frac R4t_\alpha \,,\\
  \n_\mu\n_\alpha\n_\beta t^\mu&=&\n_\alpha\n_\beta\n_\mu t^\mu+\frac
R4(\n_\alpha t_\beta+\n_\beta t_\alpha)+\frac R{12}(\n_\alpha
t_\beta-g_{\alpha\beta}\n^\mu t_\mu)\,,\\
  \n_\mu\Box t^\mu&=&\left(\Box+\frac R4\right)\n_\mu t^\mu\,,
  \label{rec2}\\
\cpd_\nu\D^\mu\n_\rho t^\sigma
&=&\cpd_\rho\D^\mu\n_\nu t^\sigma \\
&+ &\left(\delta^\sigma_\nu(\n_\rho t^\mu+\n^\mu t_\rho)-g_{
\mu\sigma }
(\n_\rho t_\nu-\n_\nu t_\rho)-\delta^\sigma_\rho(\n_\nu t^\mu+\n^\mu
t_\nu)+\delta^\mu_\nu\n_\rho t
^\sigma-\delta^\mu_\rho\n_\nu t^\sigma\right)\,.  \nonumber
\end{eqnarray}
Therefore, for a transverse vector $\n_\mu A^\mu=0$, one obtains
\begin{eqnarray}
  \n_\mu\n_\alpha A^\mu&=&\frac R4A_\alpha \,,\\
  \n_\mu\n_\alpha\n_\beta A^\mu&=&\frac
R4(\n_\alpha A_\beta+\n_\beta A_\alpha)+\frac R{12}\n_\alpha
A_\beta \,, \\
  \n_\mu\Box A^\mu&=&0\,.
\end{eqnarray}
The very last formula tells us that $\Box A_\mu$ is also a transverse vector.

Given an arbitrary (symmetric) tensor $t_{\mu\nu}$ we have
\begin{eqnarray}
  \n_\mu\n_\alpha t^{\mu\nu}&=&\n_\alpha\n_\mu t^{\mu\nu}+\frac
R3t^\nu_\alpha-\frac R{12}t^\mu_\mu\delta^\nu_\alpha \,,\\
  \n_\mu\n_\alpha\n_\beta t^{\mu\nu}&=&\n_\alpha\n_\beta\n_\mu t^{\mu\nu}+\frac
{5R}{12}\n_\alpha t^\nu_\beta+\frac R3\n_\beta t^\nu_\alpha-\frac
R{12}(\delta^\nu_\beta\n_\alpha +\delta^\nu_\alpha\n_\beta)t^\mu_\mu-\frac
R{12}g_{\alpha\beta}\n_\mu t^{\mu\nu}\,, \nonumber\\\\
  \n_\mu\Box t^{\mu\nu}&=&\left(\Box+\frac {5R}{12}\right)\n_\mu
t^{\mu\nu}-\frac R6\n^\nu t^\mu_\mu \,,\\
  \n^\sigma\n_\rho\n_\sigma t_{\mu\nu}&=&\Box\n_\rho t_{\mu\nu}+\frac
{R}{12}\left(g_{\mu\rho}\n_\sigma t^\sigma_\nu+g_{\nu\rho}\n_\sigma
t^\sigma_\mu\right)-\frac R{12}\left(\n_\mu t_{\rho\nu}+\n_\nu
t_{\rho\mu}\right)\,, \\
&=&\n_\rho\Box t_{\mu\nu}-\frac
{R}{12}\left(g_{\mu\rho}\n_\sigma t^\sigma_\nu+g_{\nu\rho}\n_\sigma
t^\sigma_\mu\right)+\frac R{12}\left(\n_\mu t_{\rho\nu}+\n_\nu
t_{\rho\mu}\right)+\frac R4\n_\rho h_{\mu\nu}.\nonumber \\
\end{eqnarray}
Therefore, for a transverse (symmetric) tensor $\n_\mu T^{\mu\nu}=0$, one similarly obtains
\begin{eqnarray}
  \n_\mu\n_\alpha T^{\mu\nu}&=&\frac
R3T^\nu_\alpha-\frac R{12}T^\mu_\mu\delta^\nu_\alpha \,, \\
  \n_\mu\n_\alpha\n_\beta T^{\mu\nu}&=&\frac
{5R}{12}\n_\alpha T^\nu_\beta+\frac R3\n_\beta T^\nu_\alpha-\frac
R{12}(\delta^\nu_\beta\n_\alpha +\delta^\nu_\alpha\n_\beta)T^\mu_\mu \,, \\
  \n_\mu\Box T^{\mu\nu}&=&-\frac R6\n^\nu T^\mu_\mu\,,\\
  \n^\sigma\n_\rho\n_\sigma T_{\mu\nu}&=&\Box\n_\rho T_{\mu\nu}-\frac
R{12}\left(\n_\mu T_{\rho\nu}+\n_\nu T_{\rho\mu}\right)\,,\\
&=&\n_\rho\Box T_{\mu\nu}+\frac R{12}\left(\n_\mu T_{\rho\nu}+\n_\nu
T_{\rho\mu}\right)+\frac R4\n_\rho T_{\mu\nu}\,.
\end{eqnarray}
Next, for a traceless (symmetric) tensor $H^{\mu}_{\mu}=0$ one gets
\begin{eqnarray}
  \n_\mu\n_\alpha H^{\mu\nu}&=&\n_\alpha\n_\mu H^{\mu\nu}+\frac
R3H^\nu_\alpha\,,\\
  \n_\mu\n_\alpha\n_\beta H^{\mu\nu}&=&\n_\alpha\n_\beta\n_\mu H^{\mu\nu}+\frac
{5R}{12}\n_\alpha H^\nu_\beta+\frac R3\n_\beta H^\nu_\alpha-\frac
R{12}g_{\alpha\beta}\n_\mu H^{\mu\nu}\,, \\
  \n_\mu\Box H^{\mu\nu}&=&\left(\Box+\frac {5R}{12}\right)\n_\mu
H^{\mu\nu}\,.\label{rec3}
\end{eqnarray}
Moreover, for a transverse and traceless (symmetric) tensor
$\n^\mu\hp_{\mu\nu}={\hp}^{\mu_\mu}=0$, one obtains:
\begin{eqnarray}
  \n_\mu\n_\alpha {\hp}^{\mu\nu}&=&\frac
R3{\hp}^\nu_\alpha \,, \\
  \n_\mu\n_\alpha\n_\beta {\hp}^{\mu\nu}&=&\frac
{5R}{12}\n_\alpha {\hp}^\nu_\beta+\frac R3\n_\beta {\hp}^\nu_\alpha \,, \\
  \n_\mu\Box {\hp}^{\mu\nu}&=&0\,.
\end{eqnarray}
The very last formula tells us that $\Box \hp_{\mu\nu}$ is also a transverse
and traceless tensor.
For completeness, we also note
\begin{equation}
\n_\rho\Box {\hp}_{\mu\nu}=\Box \n_\rho{\hp}_{\mu\nu}-\frac R{6}\left(\n_\mu
{\hp}_{\rho\nu}+\n_\nu
{\hp}_{\rho\mu}\right)-\frac R4\n_\rho {\hp}_{\mu\nu}\,.
\end{equation}
From all the above three recursion relations, we can deduce a simple relation. For a scalar (from
Eq.~(\ref{rec1}))
\begin{eqnarray}
  \n_\mu\Box \varphi&=&\left(\Box-\frac R4\right)\n_\mu \varphi~\Rightarrow~
  \n_\mu\Box^n \varphi=\left(\Box-\frac R4\right)^n\n_\mu \varphi\,.
\end{eqnarray}
For any arbitrary vector, no transversality is required (from (\ref{rec2}))
\begin{eqnarray}
  \n_\mu\Box t^\mu&=&\left(\Box+\frac R4\right)\n_\mu t^\mu~\Rightarrow~
  \n_\mu\Box^n t^\mu=\left(\Box+\frac R4\right)^n\n_\mu t^\mu\,.
\end{eqnarray}
For a traceless (symmetric) and not necessarily transverse tensor (from
(\ref{rec3}))
\begin{eqnarray}
  \n_\mu\Box H^{\mu\nu}&=&\left(\Box+\frac {5R}{12}\right)\n_\mu
H^{\mu\nu}~\Rightarrow~
  \n_\mu\Box^n H^{\mu\nu}=\left(\Box+\frac {5R}{12}\right)^n\n_\mu
H^{\mu\nu}\,.
\end{eqnarray}

Two more extremely essential commutators are needed. The first is for a 3-rank
tensor
$t^{\beta\mu\alpha}$. One can compute the following relation
\begin{equation}
\begin{split}
\n_\alpha\Box t^{\beta\mu\alpha}&=\left(\Box+\frac R4\right)\n_\alpha
t^{\beta\mu\alpha}+\frac R6\n_\rho (t^{\rho\mu\beta}+t^{\beta\rho\mu})
-\frac R6(\n^\beta
t_\alpha^{\phantom{\alpha}\mu\alpha}+\n^\mu
t_{\beta\alpha}^{\phantom{\beta\alpha
} \alpha } )\,.
\end{split}
\end{equation}
We note that on the RHS, the second term is a specific linear
combination of the initial tensor with some index permutations, while the last
piece is a combination of various traces.
Given a tensor $V^{\beta\mu\alpha}$, which enjoys the following properties
$$
V^{\beta\mu\alpha}+V^{\mu\alpha\beta}+V^{\alpha\beta\mu}=0,~V^{\beta\mu\alpha}
=V^{\mu\beta\alpha},~V^{\alpha}_{\phantom{\alpha}\mu\alpha}=V^{\mu\alpha}_{
\phantom {\alpha\mu}\alpha}=0\,,
$$
one comes to a simple relation
\begin{equation}
\begin{split}
\n_\alpha\Box V^{\beta\mu\alpha}&=\left(\Box+\frac R{12}\right)\n_\alpha
V^{\beta\mu\alpha}\,.
\end{split}\label{rec33}
\end{equation}
Notice that outlined symmetry properties are very much similar (not identical
though) to those of the so called {\it Cotton tensor}.
A recursion relation following from the latter formula, reads
\begin{equation}
\begin{split}
\n_\alpha\Box^n V^{\beta\mu\alpha}&=\left(\Box+\frac R{12}\right)^n\n_\alpha
V^{\beta\mu\alpha}\,.
\end{split}
\end{equation}

The last relation we need is for a 4-rank
tensor
$t^{\mu\alpha\nu\beta}$. One can compute the following relation
\begin{equation}
\begin{split}
\n_\mu\Box t^{\mu\alpha\nu\beta}&=\left(\Box+\frac R4\right)\n_\mu
t^{\mu\alpha\nu\beta}+\frac R6\n_\rho
\left(t^{\alpha\rho\nu\beta}+t^{\nu\alpha\rho\beta}+t^{\beta\alpha\nu\rho}
\right)\\
&-\frac R6\left(\n^\alpha t^{\mu\phantom{\mu}\nu\beta}_{\phantom{\mu}\mu}
+\n^\nu t^{\mu\alpha\phantom{\mu}\beta}_{\phantom{\mu\alpha}\mu}
+\n^\beta t^{\mu\alpha\nu}_{\phantom{\mu\alpha\nu}\mu}\right)\,.
\end{split}
\end{equation}
One immediately sees that on the RHS, the second term is a specific linear
combination of the initial tensor with some index permutations while the last
piece is a combination of various traces.
Given a tensor $W^{\mu\alpha\nu\beta}$, which enjoys all the symmetric
properties of the Weyl tensor, and also totally traceless, one comes to a simple
relation
\begin{equation}
\begin{split}
\n_\mu\Box W^{\mu\alpha\nu\beta}&=\left(\Box+\frac R4\right)\n_\mu
W^{\mu\alpha\nu\beta}\,.
\end{split}\label{rec44}
\end{equation}
A recursion relation following from the latter formula, reads
\begin{equation}
\begin{split}
\n_\mu\Box^n W^{\mu\alpha\nu\beta}&=\left(\Box+\frac R4\right)^n\n_\mu
W^{\mu\alpha\nu\beta}\,.
\end{split}
\end{equation}
\section{Cancellation of Modes}
\subsection{vector mode}\label{apA}

In this section all the derivative operators, metric and curvatures take their background values, and we omit the bars.
For the corresponding piece of $
r^{\mu\sigma}_{\phantom{\mu}\nu\rho}$, we have
\begin{equation*}
\begin{split}
{r}^{\sigma\mu}_ { \phantom{\mu}\nu\rho }(\Ap_{\mu})&=
\frac R{24}(\delta^\mu_\nu
(\n^\sigma\Ap_\rho+\n_\rho{\Ap}^\sigma)-\delta^\mu_\rho
(\n^\sigma\Ap_\nu+\n_\nu{\Ap}^\sigma)-\delta^\sigma_\nu
(\n^\mu\Ap_\rho+\n_\rho{\Ap}^\mu)+\\
\delta^\sigma_\rho
&(\n^\mu\Ap_\nu+\n_\nu{\Ap}^\mu))
+\frac12
\left(\cpd_\nu\D^\mu
(\n^\sigma\Ap_\rho+\n_\rho{\Ap}^\sigma)-\D_\nu\D^\sigma
(\n^\mu\Ap_\rho+\n_\rho{\Ap}^\mu)\right.\\
&\left.-\cpd_\rho\D^\mu
(\n^\sigma\Ap_\nu+\n_\nu{\Ap}^\sigma)+\D_\rho\D^\sigma
(\n^\mu\Ap_\nu+\n_\nu{\Ap}^\mu)
\right)\,.
\end{split}
\end{equation*}
Now we do the following commutations in the two last lines. In the first term,
with ${\Ap}_\rho$ and the first term with ${\Ap}_\nu$, we
commute $\sigma$ and $\mu$ derivatives.
In the second line for the terms with
${\Ap}^\sigma$ and ${\Ap}^\mu$, we exchange $\rho$ and $\nu$ derivatives. After
some algebra together with the Riemann tensor substitution, we get
\begin{equation*}
\begin{split}
&\frac{24}R{r}^{\sigma\mu}_ { \phantom{\mu}\nu\rho }(\Ap_{\mu})=
\delta^\mu_\nu
(\n^\sigma\Ap_\rho+\n_\rho{\Ap}^\sigma)-\delta^\mu_\rho
(\n^\sigma\Ap_\nu+\n_\nu{\Ap}^\sigma)-\delta^\sigma_\nu
(\n^\mu\Ap_\rho+\n_\rho{\Ap}^\mu)\\
&+\delta^\sigma_\rho
(\n^\mu\Ap_\nu+\n_\nu{\Ap}^\mu)
+\left(\delta_\rho^\mu\n_\nu{\Ap}^\sigma-\delta^\sigma_\rho\n_\nu{\Ap}
^\mu\right)\\
&+\left(\delta^\sigma_\nu(\n_\rho{\Ap}^\mu+\n^\mu{\Ap}_\rho)-g_{\mu\sigma}
(\n_\rho{\Ap}_\nu-\n_\nu{\Ap}_\rho)-\delta^\sigma_\rho(\n_\nu{\Ap}^\mu+\n^\mu{
\Ap}_\nu)+\delta^\mu_\nu\n_\rho{\Ap}^\sigma-\delta^\mu_\rho\n_\nu{\Ap}
^\sigma\right)\\
&-\left(\delta^\mu_\nu(\n_\rho{\Ap}^\sigma+\n^\sigma{\Ap}_\rho)-g_{\mu\sigma}
(\n_\rho{\Ap}_\nu-\n_\nu{\Ap}_\rho)-\delta^\mu_\rho(\n_\nu{\Ap}^\sigma+\n^\sigma
{
\Ap}_\nu)+\delta^\sigma_\nu\n_\rho{\Ap}^\mu-\delta^\sigma_\rho\n_\nu{\Ap}
^\mu\right)\\
&-\left(\delta_\nu^\mu\n_\rho{\Ap}^\sigma-\delta^\sigma_\nu\n_\rho{\Ap}
^\mu\right)=0\,.
\end{split}
\end{equation*}
Note that all the terms cancels explicitly.

Since $r^{\sigma}_{\rho}$ and $r$ are obtained by a simple contraction of $
r^{\mu\sigma}_{\phantom{\mu}\nu\rho}$ with the Kronecker delta our result
implies that the piece ${\Ap}_\nu$ is also absent in $r$ and $
r^{\sigma}_{\rho}$. Similarly,
\begin{equation}
\begin{split}
\delta_0({\Ap}_\mu)&=\frac14 (\n_\mu\Ap_\nu+\n_\nu{\Ap}_\mu)\Box
(\n^\mu{\Ap}^\nu+\n^\nu{\Ap}^\mu)\\
&+\frac12\D_\mu (\n^\mu{\Ap}^\rho+\n^\rho{\Ap}^\mu)\D_\nu
(\n^\nu{\Ap}_\rho+\n_\rho{\Ap}^\nu)\\
&-\frac R{24}(\n_\mu\Ap_\nu+\n_\nu{\Ap}_\mu)
(\n^\mu{\Ap}^\nu+\n^\nu{\Ap}^\mu)\,.
\end{split}
\end{equation}
Note that $\delta_0$ is an integrand, and we can integrate it by parts. Doing so in the first and last lines, and utilising several
commutation relations, we get
\begin{equation}
\begin{split}
\delta_0({\Ap}_\mu)&=-\frac12
\Ap_\nu\left(\Box+\frac{5R}{12}\right)\n_\mu
(\n^\mu{\Ap}^\nu+\n^\nu{\Ap}^\mu)+\frac12{\Ap}^\rho\left(\Box+\frac
R4\right)^2{\Ap}_\rho+\frac R{12}{\Ap}^\rho\left(\Box+\frac
R4\right){\Ap}_\rho\,, \\
&={\Ap}^\rho\left[-\frac12
\left(\Box+\frac{5R}{12}\right)+
\frac12\left(\Box+\frac
R4\right)+\frac R{12}\right]\left(\Box+\frac
R4\right){\Ap}_\rho=0\,.
\end{split}
\end{equation}

\subsection{Scalar Mode, $\n_\mu\n_\nu B$}\label{apB}

As in previous section, all the derivative operators, metric and curvatures take their background values, and we omit the bars.
For the corresponding piece of $
r^{\mu\sigma}_{\phantom{\mu}\nu\rho}$, we have
\begin{equation}
\begin{split}
2
r^{\mu\sigma}_{\phantom{\mu}\nu\rho}(\n_\mu\n_\nu
B)&=(\n_\nu\n^\mu\n^\sigma\n_\rho
-\n_\nu\n^\sigma\n^\mu\n_\rho
-\n_\rho\n^\mu\n^\sigma\n_\nu
+\n_\rho\n^\sigma\n^\mu\n_\nu)B\\
&+2\frac R{24}(\delta^\mu_\nu\n^\sigma\n_\rho-\delta^\mu_\rho\n^\sigma\n_\nu
-\delta^\sigma_\nu\n^\mu\n_\rho+\delta^\sigma_\rho\n^\mu\n_\nu)
B\,.
\end{split}
\end{equation}
Since $B$ is a scalar, the two most right derivatives can always be commuted.
Also, we can commute other in order to cancel explicit 4-derivative terms.
Explicitly, we can commute $\sigma$ and $\mu$ derivatives in the first and last
terms in the first line. Doing
so, together with the Riemann tensor substitution, we gain four
2-derivative terms as follows:
\begin{equation}
\begin{split}
2\frac{12}R
r^{\mu\sigma}_{\phantom{\mu}\nu\rho}(\n_\mu\n_\nu
B)&=(
\n_\nu
(\delta^\mu_\rho\n^\sigma-\delta^\sigma_\rho\n^\mu)
+\n_\rho(\delta^\sigma_\nu\n^\mu-\delta^\mu_\nu\n^\sigma)
)B\\
&+(\delta^\mu_\nu\n^\sigma\n_\rho-\delta^\mu_\rho\n^\sigma\n_\nu
-\delta^\sigma_\nu\n^\mu\n_\rho+\delta^\sigma_\rho\n^\mu\n_\nu)
B=0\,.
\end{split}
\end{equation}
Now we have only two derivatives everywhere acting on a
scalar. We therefore can forget ordering those derivatives.
An explicit cancellation of all terms is transparent.

Since $
r^{\sigma}_{\rho}$ and $r$ are obtained by a simple contraction of $
r^{\mu\sigma}_{\phantom{\mu}\nu\rho}$ with the Kronecker delta, our result
implies that the piece $\n_\mu\n_\nu B$ is also absent in $r$ and $
r^{\sigma}_{\rho}$.

Now,
\begin{equation}
\begin{split}
\delta_0(\n_\mu\n_\nu
B)&=B\left(\frac14\n_\mu\n_\nu\Box\n^\mu\n^\nu-\frac14\Box^3
+\frac12\Box\n_\nu\n_\rho\n^\nu\n^\rho-\frac12\n_\mu\n^\rho\n^\mu\n_\nu\n_\rho\n
^\nu\right)B\\
&-\frac R{48}B(\Box^2+2\n^\nu\n_\mu\n_\nu\n^\mu)B\,,
\end{split}
\end{equation}
where we implicitly used the fact that the actual computation goes under the
integral. As a result we can imploy integration by parts.
Performing the first iteration of commutations, one yields
\begin{equation}
\begin{split}
\delta_0(\n_\mu\n_\nu
B)&=B\left(\frac14\n_\mu\left(\left(\Box+\frac5{
12}R\right)\Box\n^\mu-\frac R6\n^\mu\Box\right)
-\frac14\Box^3
+\frac12\Box^2\left(\Box+\frac
R4\right)\right.\\
&\left.-\frac12\n_\mu\n^\rho\n^\mu\n_\rho\left(\Box+\frac
R4\right)\right)B
-\frac R{48}B\left(\Box^2+2\Box\left(\Box+\frac
R4\right)\right)B
\end{split}
\end{equation}
Performing the remaining possible commutations, one finally gets
\begin{equation}
\begin{split}
\delta_0(\n_\mu\n_\nu
B)&=B\left(\frac14\left(\left(\Box+\frac2{
3}R\right)\left(\Box+\frac R4\right)\Box-\frac R6\Box^2\right)
-\frac14\Box^3
+\frac12\Box^2\left(\Box+\frac
R4\right)-\frac12\Box\left(\Box+\frac
R4\right)^2\right)B\\
&-\frac R{48}B\left(\Box^2+2\Box\left(\Box+\frac
R4\right)\right)B=0\,.
\end{split}
\end{equation}

\end{document}